\documentclass[twocolumn]{aastex631}

\usepackage{amsmath}
\usepackage{amssymb}
\usepackage{hyperref}
\usepackage{natbib}

\newcommand{\Teff}{T_{\rm eff}}
\newcommand{\Acc}{A_{\text{cc}}}
\newcommand{\AIa}{A_{\text{Ia}}}

\defcitealias{Weinberg2022}{W22}

\begin{document}

\title{Chemical Cartography with APOGEE: Two-process Parameters and Residual Abundances for 288,789 Stars from Data Release 17}

\correspondingauthor{Tawny Sit}
\email{sit.6@osu.edu}

\author[0000-0001-8208-9755]{Tawny Sit}
\affiliation{Department of Astronomy, The Ohio State University, Columbus, OH 43210, USA}

\author[0000-0001-7775-7261]{David H. Weinberg}
\affiliation{Department of Astronomy, The Ohio State University, Columbus, OH 43210, USA}

\author[0000-0001-7339-5136]{Adam Wheeler}
\affiliation{Department of Astronomy, The Ohio State University, Columbus, OH 43210, USA}

\author[0000-0003-2969-2445]{Christian R. Hayes}
\affiliation{NRC Herzberg Astronomy and Astrophysics, 5071 West Saanich Road, Victoria, BC V9E 2E7, Canada}
\affiliation{Space Telescope Science Institute, 3700 San Martin Drive, Baltimore, MD 21218, USA}

\author[0000-0001-5388-0994]{Sten Hasselquist}
\affiliation{Space Telescope Science Institute, 3700 San Martin Drive, Baltimore, MD 21218, USA}

\author[0000-0002-6939-0831]{Thomas Masseron}
\affiliation{Instituto de Astrof{\'i}sica de Canarias, C/Via L{\'a}ctea s/n, E-38205 La Laguna, Tenerife, Spain}
\affiliation{Departamento de Astrof{\'i}sica, Universidad de La Laguna, E-38206 La Laguna, Tenerife, Spain}

\author[0000-0002-4989-0353]{Jennifer Sobeck}
\affiliation{IPAC, MC 314-6, Caltech, 1200 E. California Blvd., Pasadena, CA 91125}

\begin{abstract}
Stellar abundance measurements are subject to systematic errors that induce extra scatter and artificial correlations in elemental abundance patterns.  We derive empirical calibration offsets to remove systematic trends with surface gravity $\log(g)$ in 17 elemental abundances of 288,789 evolved stars from the SDSS APOGEE survey.  We fit these corrected abundances as the sum of a prompt process tracing core-collapse supernovae and a delayed process tracing Type Ia supernovae, thus recasting each star's measurements into the amplitudes $\Acc$ and $\AIa$ and the element-by-element residuals from this two-parameter fit.  As a first application of this catalog, which is $8\times$ larger than that of previous analyses that used a restricted $\log(g)$ range, we examine the median residual abundances of 14 open clusters, nine globular clusters, and four dwarf satellite galaxies.  Relative to field Milky Way disk stars, the open clusters younger than 2 Gyr show $\approx 0.1-0.2$ dex enhancements of the neutron-capture element Ce, and the two clusters younger than 0.5 Gyr also show elevated levels of C+N, Na, S, and Cu.  Globular clusters show elevated median abundances of C+N, Na, Al, and Ce, and correlated abundance residuals that follow previously known trends.  The four dwarf satellites show similar residual abundance patterns despite their different star formation histories, with $\approx 0.2-0.3$ dex depletions in C+N, Na, and Al and $\approx 0.1$ dex depletions in Ni, V, Mn, and Co.  We provide our catalog of corrected APOGEE abundances, two-process amplitudes, and residual abundances, which will be valuable for future studies of abundance patterns in different stellar populations and of additional enrichment processes that affect galactic chemical evolution. 
\end{abstract}

\section{Introduction} \label{sec:intro}

With highly multiplexed, high-resolution spectra, the APOGEE \citep{Majewski2017} and GALAH \citep{DeSilva2015} surveys have measured detailed chemical fingerprints of hundreds of thousands of stars in the Milky Way (MW) and its closest satellites.  Although these surveys measure 15-20 elemental abundances in each star, the abundance patterns of most MW disk and bulge stars can be predicted to surprisingly high accuracy by a two-dimensional fit, based on, e.g., the Mg and Fe abundances or Fe abundance and age \citep{Weinberg2019,Weinberg2022,Griffith2019,Griffith2021,Griffith2022,Griffith2023,Ness2019,Ness2022,Ting2022,Ratcliffe2023}.  For well measured elements, the rms intrinsic scatter of residuals from these two-parameter fits is typically 0.01-0.04 dex.  While this scatter is comparable to the statistical measurement errors for individual stars, the correlation of residuals demonstrates rich underlying structure that encodes information about the astrophysical sources of the elements and the processes that govern chemical enrichment \citep{Ting2022,Griffith2022,Weinberg2022}.

Abundance measurements in any spectroscopic survey are subject to systematic errors caused by imperfections in the model atmospheres, spectral synthesis codes, and analysis pipelines used to infer the abundances from the spectra.  Because the evolved stars targeted by APOGEE span a wide range of $\log(g)$ and $\Teff$, {\it differential} systematics across the sample can artificially inflate the scatter of residual abundances and produce spurious correlations of residuals. Previous studies have addressed this problem by restricting the $\log(g)$ and $\Teff$ range of the sample \citep{Ting2022,Weinberg2022}, by resampling populations to a matched $\log(g)$ distribution before comparing abundance trends \citep{Griffith2021}, or by applying a local calibration fit to nearest neighbors in a parameter space that includes $\log(g)$ \citep{Ness2022}. \citet{Eilers2022} fit an abundance model depending on orbital actions and $\log(g)$, then subtract off the $\log(g)$ dependencies, in essence making the assumption that all stars on the same orbit have the same birth abundances. In this paper we use an empirical approach to calibrate the systematic trends of APOGEE abundances with $\log(g)$, which enables us to derive corrected abundances that greatly suppress differential systematics across the range $0 \leq \log(g) \leq 3.5$.  \cite{Weinberg2019} and \cite{Griffith2019} show that the median trends of [X/Mg] vs. [Mg/H] are nearly universal throughout the MW disk and bulge for all well measured APOGEE elements, provided one separates the low-$\alpha$ and high-$\alpha$ stellar populations. Given this universality, and the fact that any individual star evolves through a wide range of $\log(g)$ during its post-main sequence lifetime, the abundances of most elements in a star remain relatively constant throughout its life. Therefore, we assume that any dependence of the median trend on $\log(g)$ is most likely a consequence of abundance systematics rather than a genuine physical effect. We make this assumption to derive calibration offsets as a function of $\log(g)$ and [Mg/H] that force all [X/Mg] median trends to match those of $\log(g)=1.75$ stars.  With a large sample, median abundances can be measured to precision much higher than that of individual stars, so we can robustly derive offsets at the 0.01-dex level.

After deriving corrected abundances, we follow the approach of Weinberg et al.\ (\citeyear{Weinberg2022}, hereafter \citetalias{Weinberg2022}) and fit each sample star with a two-process model \citep{Weinberg2019,Weinberg2022,Griffith2019,Griffith2021,Griffith2022,Griffith2023} that describes the abundances as the sum of a prompt enrichment process associated with core collapse supernovae (CCSN) and a delayed enrichment process associated with Type Ia supernovae (SNIa).  This procedure recasts each star's abundances into two amplitudes $\Acc$ and $\AIa$, which capture the main enrichment trends, and element-by-element residuals, which capture the deviations from these main trends.  While we use the same APOGEE DR17 \citep{Abdurrouf2022} data set as \citetalias{Weinberg2022}, our expanded $\log(g)$ range gives us a final catalog of residual abundances with $\approx 8$ times more stars.  The expanded sample is especially valuable for stellar populations whose APOGEE targets lie mostly outside the $\log(g)$ and $\Teff$ range of \citetalias{Weinberg2022}, such as luminous low-gravity giants in dwarf satellites and the MW bulge, and red clump stars in open clusters.

As a first application of our expanded catalog, we examine the median residual abundances of 14 open clusters, nine globular clusters, and four dwarf satellites (the Large and Small Magellanic Clouds, the Sagittarius dwarf and stream, and Gaia Sausage/Enceladus).  Taking advantage of medians for large samples allows us to robustly measure 0.05-dex level abundance differences between these stellar populations and MW field stars matched in overall CCSN and SNIa enrichment.  The open cluster sample allows us to examine trends of individual elements with cluster age, similar to the analysis presented by \cite{Griffith2022} using GALAH data.  The most important product of this paper is our catalog of two-process parameters and residual abundances, which can also be used to obtain [X/H] values corrected for $\log(g)$ systematics.  We anticipate that this catalog will be useful for many investigations, including: characterizing additional enrichment sources such as AGB stars or rare supernovae; quantifying stochastic effects such as incomplete sampling of the initial mass function (IMF); revealing subtle differences in the abundance patterns of the bulge, radial and vertical zones of the disk, halo, and satellites; identifying groups with distinctive abundance patterns that could indicate a common birth environment; uncovering trends with stellar age or binarity; defining chemically homogeneous samples for reconstructing the Galactic potential \citep{Price-Whelan2021}; and identifying rare outliers that could be signs of exotic physical processes or unusual measurement errors.

This paper is organized as follows. In Section \ref{sec:data}, we describe our data and sample selection criteria from APOGEE DR17. In section \ref{sec:logg}, we describe our derivation of calibration offsets that remove systematic trends with $\log(g)$ in median abundance trends, and apply these offsets to our data. We fit the two-process model to our sample and derive residual abundances in Section \ref{sec:2proc}.  We then apply our expanded residual abundance catalog to examine residual abundances in open clusters (Section \ref{subsec:OCs}), globular clusters (Section \ref{subsec:GCs}), and MW satellite galaxies (Section \ref{subsec:gals}). Finally, we discuss and summarize our results, and outline future prospects, in Section \ref{sec:conclusions}.  Appendix~\ref{appx:catalog} provides a user's guide to the residual abundance catalog. We note that this catalog is also valuable for ``traditional" abundance analyses of APOGEE populations that span a wide range of $\log(g)$, as the [X/H] abundances themselves are corrected for $\log(g)$ systematics.

\section{Data} \label{sec:data}

We use data from DR17 \citep{Abdurrouf2022} of the SDSS-IV \citep{Blanton2017} APOGEE survey \citep{Majewski2017}. This data release contains the completed collection of over 650,000 high-resolution ($R\sim22,500$), near-infrared (H-band, 1.51–1.70 $\mu$m) spectra from the APOGEE spectrographs \citep{Wilson2019} on the 2.5 m Sloan Foundation telescope \citep{Gunn2006} at Apache Point Observatory, New Mexico, and the du Pont telescope \citep{Bowen1973} at Las Campanas Observatory, Chile. Targeting for the APOGEE survey is described in \citet{Zasowski2013,Zasowski2017,Beaton2021,Santana2021}. Spectral data reduction and calibrations for APOGEE are performed by a dedicated data processing pipeline described in \citet{Nidever2015}; the reduced spectra then serve as inputs to the APOGEE Stellar Parameters and Abundances Pipeline (ASPCAP; \citealt{Holtzman2015,GarcíaPérez2016}).

We primarily use the stellar parameters and abundances derived by ASPCAP. ASPCAP uses FERRE \citep{AllendePrieto2006} to fit APOGEE spectra to a grid of synthetic spectra \citep{Meszaros2012,Zamora2015} generated with Synspec \citep{Hubeny2011} from MARCS model atmospheres \citep{Gustafsson2008} and an H-band line list \citep{Shetrone2015,Hasselquist2016,Cunha2017,Smith2021}, with NLTE calculations used for Na, Mg, K, and Ca \citep{Osorio2020}. Alternative analyses using spectral grids synthesized with Turbospectrum \citep{turbospec} are available, and the major relevant difference between the grids is that Turbospectrum uses 3D spherical geometry for radiative transfer in $\log(g)\leq3.0$ giants but no NLTE calculations. We choose to use the default DR17 analysis with Synspec for this study. ASPCAP uses a two-step fitting process: stellar atmospheric parameters are first determined with the full APOGEE spectrum, and then these parameters are held constant while individual abundances are fit for in smaller wavelength windows \citep{GarcíaPérez2016}. We  use 17 species measured by ASPCAP in this study: C, N, O, Na, Mg, Al, Si, S, K, Ca, V, Cr, Mn, Fe, Co, Ni, and Ce. We exclude Ti and Ti II because they show trends that deviate from those in the literature (Holtzman et al., in preparation). We use the C abundance measured from molecular lines because it was shown to be more reliable than the abundance measured from neutral atomic lines in DR16 \citep{Jonsson2020}.

Additionally, we include abundance measurements from the BACCHUS Analysis of Weak Lines in APOGEE Spectra (BAWLAS) catalog \citep{Hayes2022}. The BAWLAS catalog provides abundance measurements of elements with faint or blended lines (Na, P, S, V, Cu, Ce, and Nd) for a sample of 126,362 high signal-to-noise (S/N$>$150) spectra of APOGEE red giants using the Brussels Automatic Code for Characterizing High accUracy Spectra (BACCHUS; \citealt{BACCHUS}). It also reports C, N, and O abundances re-derived during the BACCHUS fitting process (for more details, see Section 4.5.1 of \citealt{Hayes2022}). We exclude Nd from this study because less than half of the stars in the sample used for deriving $\log(g)$ calibrations and two-process vectors (see Section \ref{subsec: sample}) had valid Nd measurements. We also exclude P because its high upper limit in [P/H] (see \citealt{Hayes2022} for more details) resulted in poor coverage over our sample's metallicity space. For the common elements between ASPCAP and BAWLAS (C, N, O, Na, S, V, and Ce), abundance measurements are treated independently, as if they were separate elements.

\citet{Weinberg2019} advocated for the use of Mg rather than Fe as a reference element because it has a single enrichment source, CCSN. We follow their example and also use Mg as the reference element in this study. Therefore, we first remove all stars whose [Fe/H] and/or [Mg/Fe] abundances are flagged so that we have a reliable $\text{[Mg/H]} = \text{[Mg/Fe]} + \text{[Fe/H]}$ value for every star in our sample. Suspect/flagged abundances are reported as Not a Number (NaN) values in the BAWLAS catalog. For consistency, we also assign NaN to BAWLAS elements for stars not in the BAWLAS sample and convert flagged measurements for all considered ASPCAP elements (other than [Fe/H] and [Mg/Fe], which were removed) into NaN. The use of NaN allows us to easily omit individual stars from calculations for specific elements, keep as many stars in the sample as possible, and simply propagate the invalid element to our final catalog.

The surface abundances of C and N vary for the red giant branch (RGB) stars that make up the majority of our sample (Section \ref{subsec: sample}) due to the CNO cycle and dredge-up events \citep{Iben1965,Shetrone2019}. However, the total amount of C+N by number remains nearly equal to the birth abundance because the extra N nuclei in the dredged-up CNO-processed material were created from C nuclei. Therefore, following the example of \citetalias{Weinberg2022} and \citet{Griffith2023}, we treat C+N as a single element, calculated as
\begin{equation}
\begin{split}
    \text{[(C+N)/H]} = & \log_{10}\left(10^{\text{[C/H]+8.39}} + 10^{\text{[N/H]}+7.78}\right) \\
    & - \log_{10}\left(10^{8.39}+10^{7.78}\right),
\end{split}
\end{equation}
where 8.39 and 7.78 are the logarithmic solar abundances for C and N, respectively, from \citet{Grevesse2007}. C+N is calculated for both the ASPCAP and BAWLAS measurements of C and N.

[X/H] abundances are calculated from the sum of [X/Fe] (reported by either ASPCAP or BAWLAS) and [Fe/H] (reported by ASPCAP). For the uncertainty on [X/H], we use the reported [X/Fe] measurement uncertainty from ASPCAP (\texttt{X\_FE\_ERR}) or the empirical [X/Fe] uncertainty from BAWLAS (\texttt{X\_FE\_ERR\_EMP}). We are primarily interested in the differential scatter between elements not between individual stars, and each star uses the same [Fe/H] measurement in our calculation of [X/H], so although [Fe/H] has its own reported ASPCAP uncertainty, the impact of the [Fe/H] uncertainty is completely correlated with [X/H], and we therefore do not consider it in our [X/H] uncertainties except in the case $\text{X}=\text{Fe}$. For the combined element C+N, we follow \citetalias{Weinberg2022} and take the [C/Fe] uncertainty as the [(C+N)/H] uncertainty. ASPCAP estimates uncertainties empirically as a function of S/N, $\Teff$, and metallicity using repeated measurements of stars, and these uncertainties typically are larger than those reported by spectral model fitting \citep[][Section 5.4]{Jonsson2020}. BAWLAS reports two sets of uncertainties: a measurement uncertainty, determined from the scatter in line-by-line measurements of a given element, and and an empirical uncertainty, estimated from repeat observations \citep[][Section 4.7]{Hayes2022}. We use the BAWLAS empirical uncertainties because they are most similar in spirit to the ASPCAP uncertainties, though there are small differences in methodology (e.g., BAWLAS empirical uncertainties do not depend on SNR or overall metallicity, but do depend on individual abundances). 

We also divide our sample (Section \ref{subsec: sample}) into high-Ia (``low-$\alpha$") and low-Ia (``high-$\alpha$") sequences following the dividing line in \citet{Weinberg2019}, where the low-Ia stars are defined as as those with:
\begin{equation}\label{eq:alpha_split}
    \begin{cases}
    \text{[Mg/Fe]} > 0.12 - 0.13\text{[Fe/H]}, & \text{[Fe/H]} < 0\\
    \text{[Mg/Fe]} > 0.12, & \text{[Fe/H]} > 0.
\end{cases}
\end{equation}
For consistency with \citet{Weinberg2019} and \citetalias{Weinberg2022}, this division is determined prior to adding the $\log(g)$ calibration and zero-point offsets (Sections \ref{sec:logg} and \ref{sec:2proc}).

\begin{deluxetable*}{cccc}[!t]
\tablecaption{Sample Sizes} \label{table:sample_sizes}
\tablehead{\colhead{} & \colhead{Low-Ia\tablenotemark{a}} & \colhead{High-Ia\tablenotemark{b}} & \colhead{Total}}
\startdata
Calibration\tablenotemark{c} & 40,513 & 111,051 & 151,564 \\
BAWLAS\tablenotemark{d} & 37,499 (18,406) & 88,662 (48,974) & 126,161 (67,380) \\
Full\tablenotemark{e} & 91,144 & 219,283 & 310,427
\enddata
\tablenotetext{a}{Sample cut defined in Equation \ref{eq:alpha_split}.}
\tablenotetext{b}{All spectra \textit{not} in the low-Ia sample.}
\tablenotetext{c}{S/N $>$ 100, \texttt{EXTRATARG = 0}, 3 kpc $\leq R \leq$ 13 kpc, $|Z|<2$ kpc, and all full sample [Mg/H], $\log(g)$, and $\Teff$ cuts. The calibration sample does not contain duplicate stars.}
\tablenotetext{d}{Spectra having at least one valid BAWLAS measurement. Parentheses indicate the number of BAWLAS stars in the calibration sample.}
\tablenotetext{e}{$-0.75\leq$ [Mg/H] $\leq0.45$, $0\leq\log(g)\leq3.5$, $3000\text{ K}\leq \Teff \leq5500\text{ K}$, and S/N $>$ 80.}
\end{deluxetable*}

\subsection{Sample Selection}\label{subsec: sample}

\begin{figure}[!t]
    \centering
    \includegraphics[width=\linewidth]{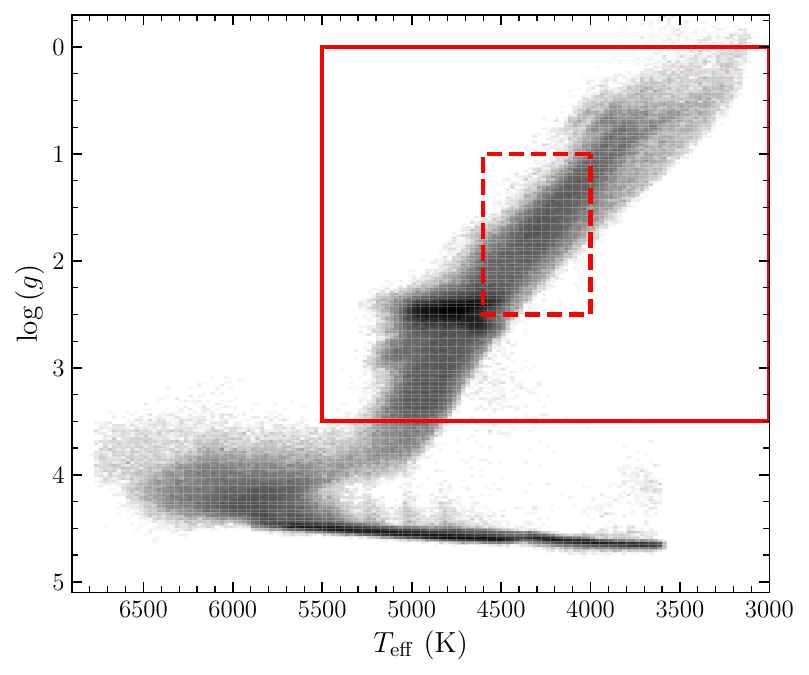}
    \caption{$\log(g)$ vs. $\Teff$ for all APOGEE DR17 stars with $-0.75\leq$ [Mg/H] $\leq0.45$ and SNR$>$80. Stars with flagged [Fe/H] and/or [Mg/Fe] are not included. We show our sample selection cuts,  $0<\log(g)<3.5$ and $3000 \text{ K}<\Teff<5500$ K, in the solid red box. For comparison, the $\log(g)$ and $\Teff$ cuts of \citetalias{Weinberg2022} are shown by the dotted red box.}
    \label{fig:teff-logg sample select}
\end{figure}

We first define a ``training" sample of DR17 stars to derive the $\log(g)$ calibrations (Section \ref{sec:logg}) and two-process vectors (Section \ref{sec:2proc}). For this calibration sample, we impose a S/N threshold of S/N $>$ 100, and we use only stars targeted as part of the main APOGEE survey by requiring the flag \texttt{EXTRATARG=0}, to avoid selection biases from special targeting classes. \citet{Weinberg2019} showed that median trends in [X/Mg]---[Mg/H] are universal throughout the Galactic disk, so we also restrict this sample to the disk by selecting stars with 3 kpc $\leq R \leq$ 13 kpc and $|Z|<2$ kpc. $R$ and $Z$ are from the AstroNN value-added catalog for DR17 \citep{LeungBovy2019a,LeungBovy2019b}, which uses APOGEE spectra and \textit{Gaia} eDR3 parallaxes to obtain spectro-photometric distances from deep learning. From \citetalias{Weinberg2022}, we require that $-0.75\leq$ [Mg/H] $\leq0.45$; this metallicity range samples a wide range of enrichment histories in the Galactic disk. Finally, we make additional cuts in $\log(g)$ and $\Teff$: $0\leq\log(g)\leq3.5$, and $3000\text{ K}\leq \Teff \leq5500\text{ K}$. These $\log(g)$ and $\Teff$ cuts select nearly the entire RGB and all red clump (RC) stars within our metallicity range, as illustrated in Figure \ref{fig:teff-logg sample select}. In total, we use 151,564 stars to derive $\log(g)$ calibrations and two-process vectors.

We use a less stringent set of cuts to define an ``application" sample of stars to which we apply our $\log(g)$ calibration, fit with a two-process model, and derive residual abundances. We require only that $-0.75\leq$ [Mg/H] $\leq0.45$, $0\leq\log(g)\leq3.5$, $3000\text{ K}\leq \Teff \leq5500\text{ K}$, and S/N $>$ 80. This larger sample includes all stars observed by APOGEE, regardless of targeting status or location in the Galaxy. These cuts result in a total of 310,427 entries in our final residual abundance catalog.

There are some stars in APOGEE that have duplicate entries (i.e., were processed separately) in ASPCAP because they were observed in multiple unique fields (see Section 2 of \citealt{Jonsson2020}). In our residual abundance catalog, each ASPCAP entry is treated as a separate star, so all duplicates (i.e., same \texttt{APOGEE\_ID}) whose ASPCAP parameters meet the sample selection criteria are included. We find that including the \texttt{FIELD} or \texttt{LOCATION\_ID} columns in our final catalog is sufficient to break degeneracies in \texttt{APOGEE\_ID}. There are no duplicate stars in the smaller calibration (``training") sample. In the larger final (``application") sample, there are 288,789 unique stars in the total 310,427 spectra/ASPCAP entries; 17,694 stars have two or more spectra.

\section{log(g) Calibrations} \label{sec:logg}

In general, the differential impact of abundance measurement systematics can be minimized by limiting a sample to a small range of $\Teff$ and $\log(g)$. However, \citetalias{Weinberg2022} showed that subtle trends in residual abundances with $\Teff$ still exist and can induce artificial correlations between elements; they correct for these trends by applying an additional, $\Teff$-dependent corrective offset to the ASPCAP abundances. Additionally, \citet{Griffith2021} showed that $\log(g)$ systematics cause noticeable changes in median ASPCAP abundance trends. Because of the correlation between $\log(g)$ and $\Teff$ on the giant branch (Figure \ref{fig:teff-logg sample select}), one can use either parameter for calibrating systematics, and we choose $\log(g)$ because of its more intuitive connection to evolutionary state.

\begin{figure}
    \centering
    \includegraphics[width=\linewidth]{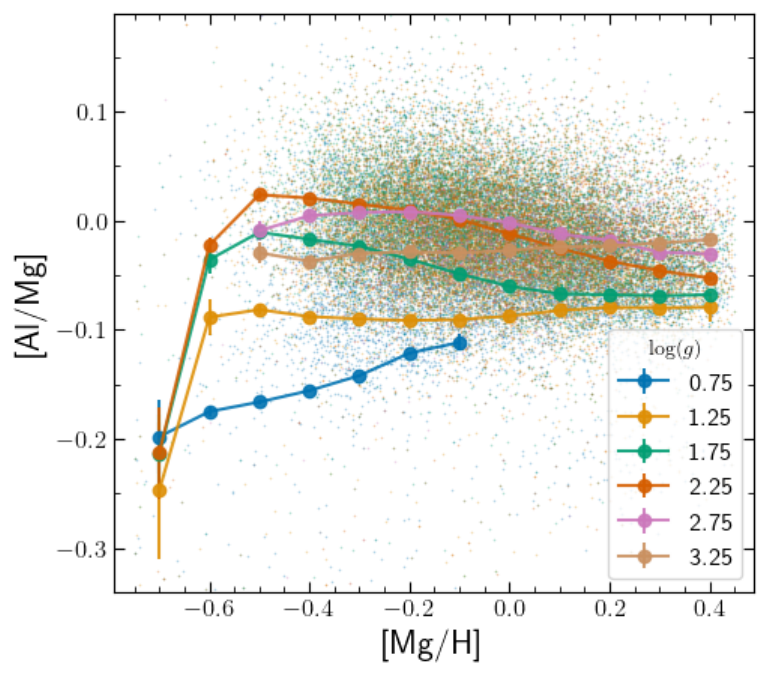}
    \caption{[Al/Mg] as a function of [Mg/H], colored by 0.5 dex bins in $\log(g)$. Stars are randomly downsampled by a factor of five to reduce crowding. In each $\log(g)$ bin, median [Al/Mg] values are calculated for every 0.1 dex [Mg/H] bin and plotted in connected large points. Error bars, usually smaller than the points, show uncertainties on the medians computed from 1000 bootstrap resamplings of each ([Mg/H], $\log(g)$) bin. Abundances are taken directly from ASPCAP with no further calibrations. This figure uses only high-Ia (low-$\alpha$) stars in the SNR$>$100 main survey disk sample used for calibration (see Section \ref{subsec: sample}), and red clump stars have been removed.}
    \label{fig:AlMg_vs_MgH}
\end{figure}

In Figure \ref{fig:AlMg_vs_MgH}, we show median abundance trends with [Mg/H] for an example element, Al, in different surface gravity bins. The stars in this figure are all high-Ia (low-$\alpha$) RGB stars in the MW disk from the main APOGEE survey sample (i.e., no special targeting). Any single star traverses a range of $\log(g)$ as it evolves off the main sequence, so we do not expect birth abundances to vary with $\log(g)$. Figure \ref{fig:AlMg_vs_MgH} shows that this is not the case for the measured Al abundance: the median [Al/Mg] vs. [Mg/H] trends are visibly different for different $\log(g)$. This figure also illustrates the power of using median abundances to identify subtle trends, since the $\log(g)$ systematics are comparable in magnitude to the star-to-star scatter but are measured at high precision in the median curves.

We remove trends with $\log(g)$ similarly to the $\Teff$ corrections in \citetalias{Weinberg2022} by applying a corrective offset to the ASPCAP or BAWLAS pipeline abundance. Our key assumption is that the median abundance trends ([X/Mg] vs. [Mg/H]) of the low-Ia and high-Ia disk populations should be independent of $\log(g)$ over the range of our sample. As discussed in the introduction, these trends are nearly universal through the disk and bulge \citep{Weinberg2019,Griffith2021},
and any individual star crosses most of our $\log(g)$ range during its red giant lifetime, so a dependence of the median trend on $\log(g)$ is much more likely to be an artifact of abundance systematics than a physical effect.

For each star, we apply offsets to the raw ASPCAP/BAWLAS abundance as follows:
\begin{equation}\label{eq:xh_corr}
    \text{[X/H]}_{\text{corr}} = \text{[X/H]}_{\text{raw}} + C^X_{\text{ZP}} + C^X_{\log(g)}(\text{[Mg/H]},\log(g)).
\end{equation}
\noindent
The zero-point offset $C_{\text{ZP}}$ is further discussed in Section \ref{sec:2proc}. The $\log(g)$ calibration correction for each star $C_{\log(g)}$ depends on the star's metallicity [Mg/H] and surface gravity $\log(g)$.

\begin{figure*}[!th]
    \centering
    \includegraphics[width=0.9\linewidth]{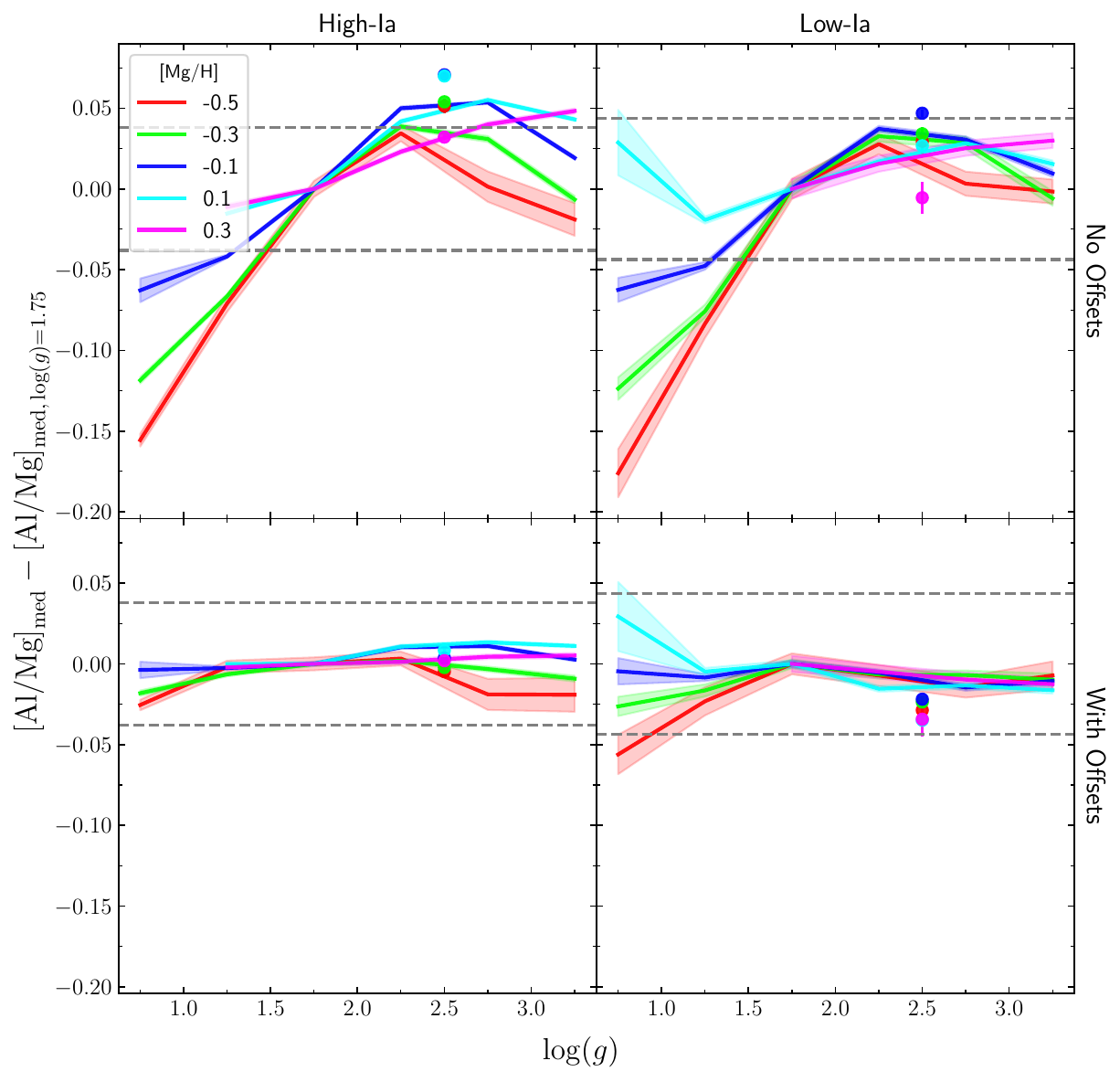}
    \caption{Median differences for Al as a function of $\log(g)$ for selected [Mg/H] bins. In each $\log(g)$ bin at a given [Mg/H], the median [Al/Mg] is computed and the median [Al/Mg] in the fiducial $\log(g)=1.75$ bin is subtracted. The median difference at $\log(g)=1.75$ is thus zero by definition for all [Mg/H]. The $\log(g)=0.25$ bins are not shown because no [Mg/H] bin at this $\log(g)$ contained more than 25 stars with valid Al measurements. The top row shows the median differences used to compute $C_{\log(g)}$, and the bottom row shows the median differences calculated after $C_{\log(g)}$ was applied. The left and right columns separate the high-Ia (low-$\alpha$) and low-Ia (high-$\alpha$) sequences, defined in Equation \ref{eq:alpha_split}; the final $C_{\log(g)}$ value is computed from the number-weighted average of the two sequences. Red, green, blue, cyan, and magenta represent $\text{[Mg/H]} = -0.5, -0.3, -0.1, 0.1, \text{and } 0.3$ bins. Filled circles represent the red clump sample. Dashed gray lines indicate the median per star measurement uncertainty for the high-/low-Ia samples. Uncertainty in the median differences is shown by the shaded regions or error bars for the red clump and is estimated from 1000 bootstrap resamplings of each ([Mg/H], $\log(g)$) bin. This figure uses only the calibration sample (Section \ref{subsec: sample}).}
    \label{fig:Al_logg_medians}
\end{figure*}

Stars in the smaller calibration sample (see Section \ref{subsec: sample}) were divided into high-Ia (low-$\alpha$) and low-Ia (high-$\alpha$) sequences following Equation \ref{eq:alpha_split}. For each sequence, we calculated the median abundances of each element in 0.1 dex bins in [Mg/H] and 0.5 dex bins in $\log(g)$. The raw ASPCAP/BAWLAS abundances were used to calculate these medians, and we required a minimum of 25 stars with valid abundance measurements in each bin to calculate a median. Red clump (RC) stars are He-core burning and thus may have slightly different systematics from the H-shell burning red giant branch (RGB) stars in the rest of our sample. Therefore, all stars in the DR17 value-added RC catalog \citep{Bovy2014} were considered as their own separate $\log(g)$ bin, and they were also split into high- and low-Ia sequences following Equation \ref{eq:alpha_split}. Then, for each [Mg/H] bin, we calculated the difference between the median [X/H] abundance in each $\log(g)$ bin from the $\log(g)$ bin centered at a fiducial value of $\log(g) = 1.75$. We plot these median differences for a subset of [Mg/H] bins for an example element, Al, in the top row of Figure \ref{fig:Al_logg_medians}. The differences are similar, though not identical, for high-Ia and low-Ia stars, supporting our interpretation that they are driven by systematics in the spectral synthesis modeling that correlate primarily with $\log(g)$ and overall metallicity.

\begin{figure*}[!htp]
    \centering
    \includegraphics[width=0.9\linewidth]{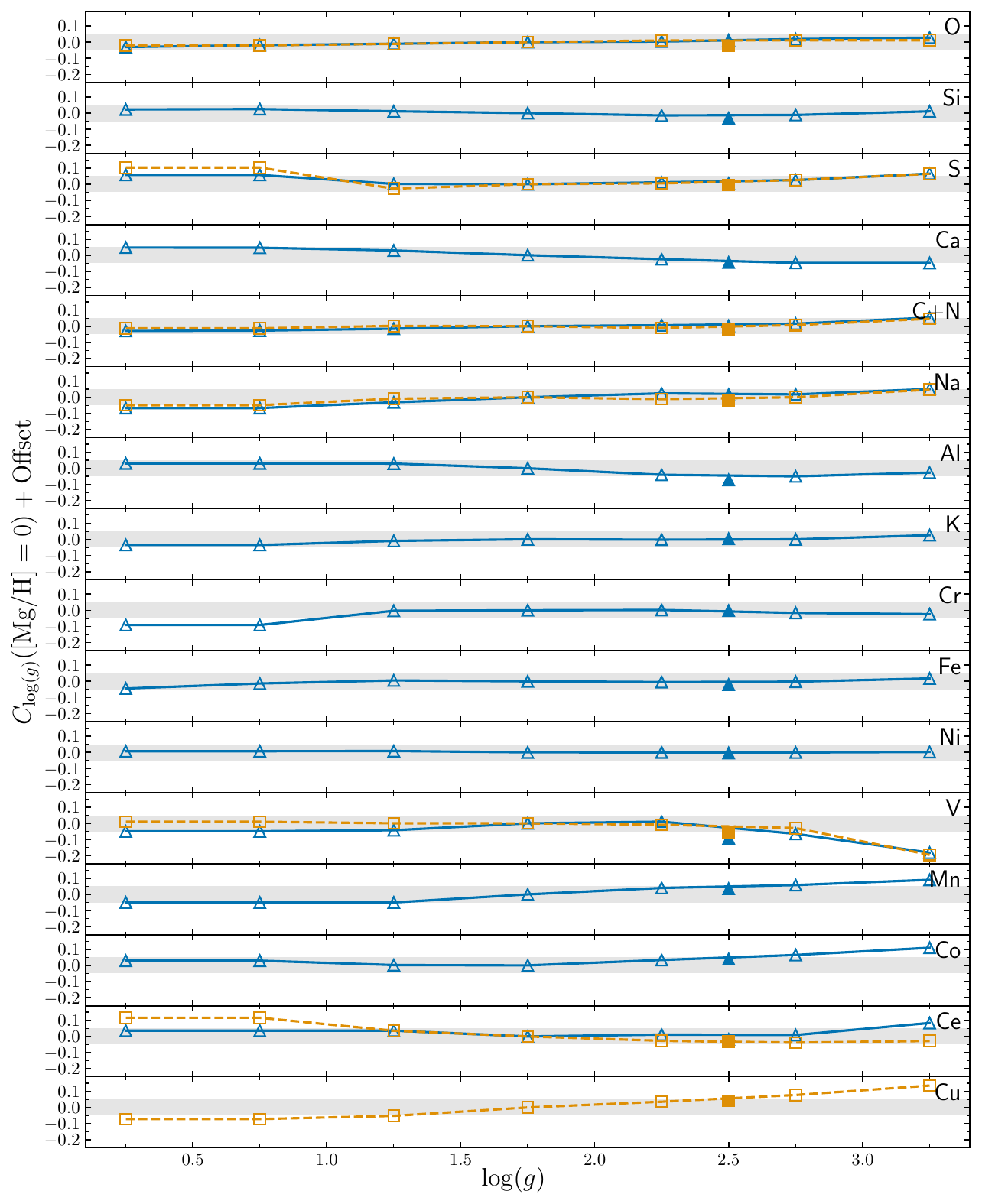}
    \caption{$\log(g)$ correction values $C^X_{\log(g)}$ as a function of $\log(g)$ for the $\text{[Mg/H]}=0$ bin. A flat offset has been applied so that each sequence passes through 0 at $\log(g)=1.75$. Blue solid lines with triangles and orange dashed lines with squares indicate ASPCAP and BAWLAS abundances, respectively. Filled triangles/squares at $\log(g)=2.5$ indicate the red clump sample. The gray band represents the $\pm0.05$ range. This figure uses the calibration sample (Section \ref{subsec: sample}).
    }
    \label{fig:logg_offsets}
\end{figure*}

Next, we took the average between the high- and low-Ia sequences of the median differences, weighted by the number of stars in the high/low-Ia bins. In doing so, we further assume that the $\log(g)$ systematics do not depend on the [$\alpha$/Fe] ratio at fixed [Mg/H]. Due to this weighted average, the median differences after calibration are not perfectly flat at $\Delta\text{[X/H]}=0$ for the high- and low-Ia sequences individually, as illustrated by the bottom row of Figure \ref{fig:Al_logg_medians}. However, in this example (and most cases) the median differences in the well-populated $\log(g)$ bins are below 0.02 dex after calibration. We could have chosen to force these differences to zero for the (larger) high-Ia sample at the expense of slightly larger deviations for the low-Ia sample. Finally, we added a flat offset to ensure that the median $\text{[X/H]}=0$ at ($\text{[Mg/H]}=0$, $\log(g)=1.75$) for the high-Ia sequence. These values define a correction function on a 2D grid of ([Mg/H], $\log(g)$) points that is the same across the high and low-Ia sequences. We plot the $C_{\log(g)}$ values as a function of $\log(g)$ at $\text{[Mg/H]}=0$ for all elements in Figure \ref{fig:logg_offsets}. For each non-RC star, we interpolate along this grid to the star's [Mg/H] and $\log(g)$ to obtain the $C_{\log(g)}$ for each element. For RC stars, we interpolate only along [Mg/H] to obtain $C_{\log(g)}$.

For most $\log(g)$ bins, the calibration offsets we apply are smaller than the typical $\approx$0.04-dex observational uncertainty in [Al/H] for individual stars. Because of the large number of stars per [Mg/H] bin, the median offset is measured much more precisely, and failing to remove these $\log(g)$ systematics would lead to spurious abundance trends. Therefore, we expect the most reliable calibrations in the most well-populated bins: $1.25\geq\log(g)\geq2.75$ and $-0.4\geq\text{[Mg/H]}\geq0.2$. Outside of this range, some elements may have too few stars in  bin ($N\leq25$) to calculate a median difference; in these cases the final calibration offset is copied from the nearest $\log(g)$ or [Mg/H] bin with a valid offset. Furthermore, even in bins with enough stars, the abundance measurements themselves can become more uncertain at the low $\log(g)$ (e.g., $\log(g)=0.25$, 0.75 bins) because the spectra of cooler stars become affected by line blends. This may contribute to the remaining spread in median differences for the low-Ia $\log(g)=0.75$ bin in Figure \ref{fig:Al_logg_medians}. However, strong trends in individual offsets outside of the overall most reliable ($\log(g)$, [Mg/H]) bins can still be fairly robust: in Figure \ref{fig:logg_offsets}, V shows the strongest $\log(g)$ correction of any element (at [Mg/H] = 0) in the $\log(g)=3.25$ bin, where the median differences used to compute the offset were calculated from 3671 stars (for ASPCAP).

\section{Two-process Model and Residual Abundances} \label{sec:2proc}

\subsection{Two-process Vectors, $q^X_{\mathrm{cc}}(z)$ and $q^X_{\mathrm{Ia}}(z)$} \label{subsec:2proc_vectors}
In this work, we adopt the two-process model as described in \citetalias{Weinberg2022}. In brief, the two-process model describes chemical abundances as a combination of IMF-averaged yields from CCSN and SNIa such that an element X produced from only these two sources will have a solar-scaled abundance described by 
\begin{equation}\label{eq:2proc}
    \text{[X/H]} = \log_{10}{\left[\Acc q^{X}_{\text{cc}}(z) + \AIa q^{X}_{\text{Ia}}(z)\right]},
\end{equation}
where the metallicity $z = 10^{\text{[Mg/H]}}$. $\Acc$ and $\AIa$ are amplitudes describing the relative amount of CCSN/SNIa contribution and are fit to each individual star. They are normalized such that $\Acc = \AIa = 1$ for a star with solar abundances, [X/H] = 0. The process vectors $q^{X}_{\text{cc}}(z)$ and $q^{X}_{\text{Ia}}(z)$ for each element X are derived from the observed [X/Mg] vs. [Mg/H] median sequences of the high-Ia and low-Ia stellar populations in the Galactic disk. The actual derivation of $q^{X}_{\text{cc}}(z)$ and $q^{X}_{\text{Ia}}(z)$ follows Section 2 of \citetalias{Weinberg2022} and makes the following key assumptions:
\begin{enumerate}
    \item Mg is a pure CCSN element, so $q^{\text{Mg}}_{\text{Ia}}(z) = 0$.
    \item The Mg and Fe processes are independent of metallicity, so $q^{\text{Mg}}_{\text{cc}}(z) = q^{\text{Mg}}_{\text{cc}}$, $q^{\text{Fe}}_{\text{cc}}(z) = q^{\text{Fe}}_{\text{cc}}$, and $q^{\text{Fe}}_{\text{Ia}}(z) = q^{\text{Fe}}_{\text{Ia}}$.
    \item The [Mg/Fe] abundance of the low-Ia population plateaus at $\text{[Mg/Fe]}_{\text{pl}}=0.3$ \citepalias[][Figure 1, left panel]{Weinberg2022}, and stars on the plateau only have Fe enrichment from CCSN ($\AIa=0$).
\end{enumerate}
\noindent
The adopted value of $\text{[Mg/Fe]}_{\text{pl}}$ is a choice, and \citet{Griffith2023} illustrate the impact of choosing slightly different values or a slightly tilted plateau.

The APOGEE abundances contain zero-point shifts chosen to make the mean abundance ratio of solar metallicity stars in the solar neighborhood [X/Fe] = 0 (\citealt{Jonsson2020}; Holtzman et al. in prep). The BAWLAS abundances are similarly calibrated to a zero-point derived from a sample of solar neighborhood stars \citep{Hayes2022}. To ensure that stars on the high-Ia/low-$\alpha$ sequence with solar metallicity have solar [X/Mg] abundances, we apply an additional flat zero-point offset $C^X_{\text{ZP}}$ that forces the median [X/Mg] trend of the high-Ia population to pass through $\text{[X/Mg]}=0$ in the $\text{[Mg/H]}=0$ bin. The spirit of our $C_{\text{ZP}}$ correction is similar to that of the APOGEE and BAWLAS analyses, but because our sample is distinct the zero-point offsets are not identical. We effectively assume that the solar abundance mixture is typical of high-Ia disk stars with [Mg/H] = 0. $C_{\text{ZP}}$ is derived from the SNR$>$100 main survey disk sample, and the abundances used to calculate the median values had the $\log(g)$ calibration $C_{\log(g)}$ previously applied so that they trace the $\log(g)=1.75$ stars. We provide a table of $C_{\text{ZP}}$ and $C_{\log(g)}$ values along with sample code for applying the calibrations in Appendix \ref{appx:calib_offsets}.

\begin{figure}[!t]
    \centering
    \includegraphics[width=\linewidth]{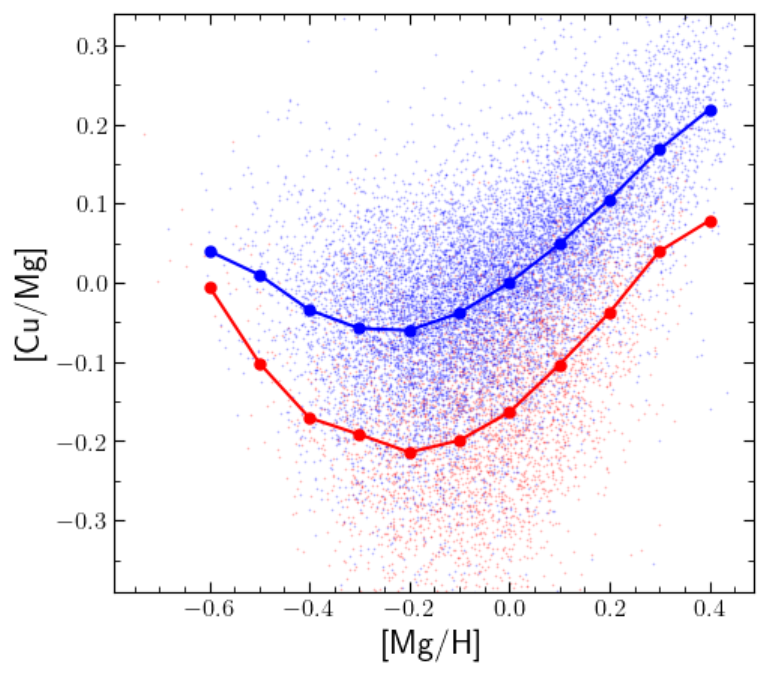}
    \caption{[Cu/Mg] vs. [Mg/H] for the SNR$>$100 main survey disk sample. High-Ia stars are in blue and low-Ia stars in red. Stars are randomly downsampled by a factor of two to reduce crowding. Connected large points indicate the median values in bins of [Mg/H]. Abundances in this plot have been corrected following Equation \ref{eq:xh_corr} so that $\log(g)$ trends are removed and the high-Ia median sequence (large blue points) passes through $\text{[Cu/Mg]}=0$ at $\text{[Mg/H]}=0$. The separation of low-Ia and high-Ia sequences indicates a significant delayed contribution to Cu, which could come from SNIa, AGB stars, or both.}
    \label{fig:CuMg}
\end{figure}

\begin{figure*}[!th]
    \centering
    \includegraphics[width=0.8\linewidth]{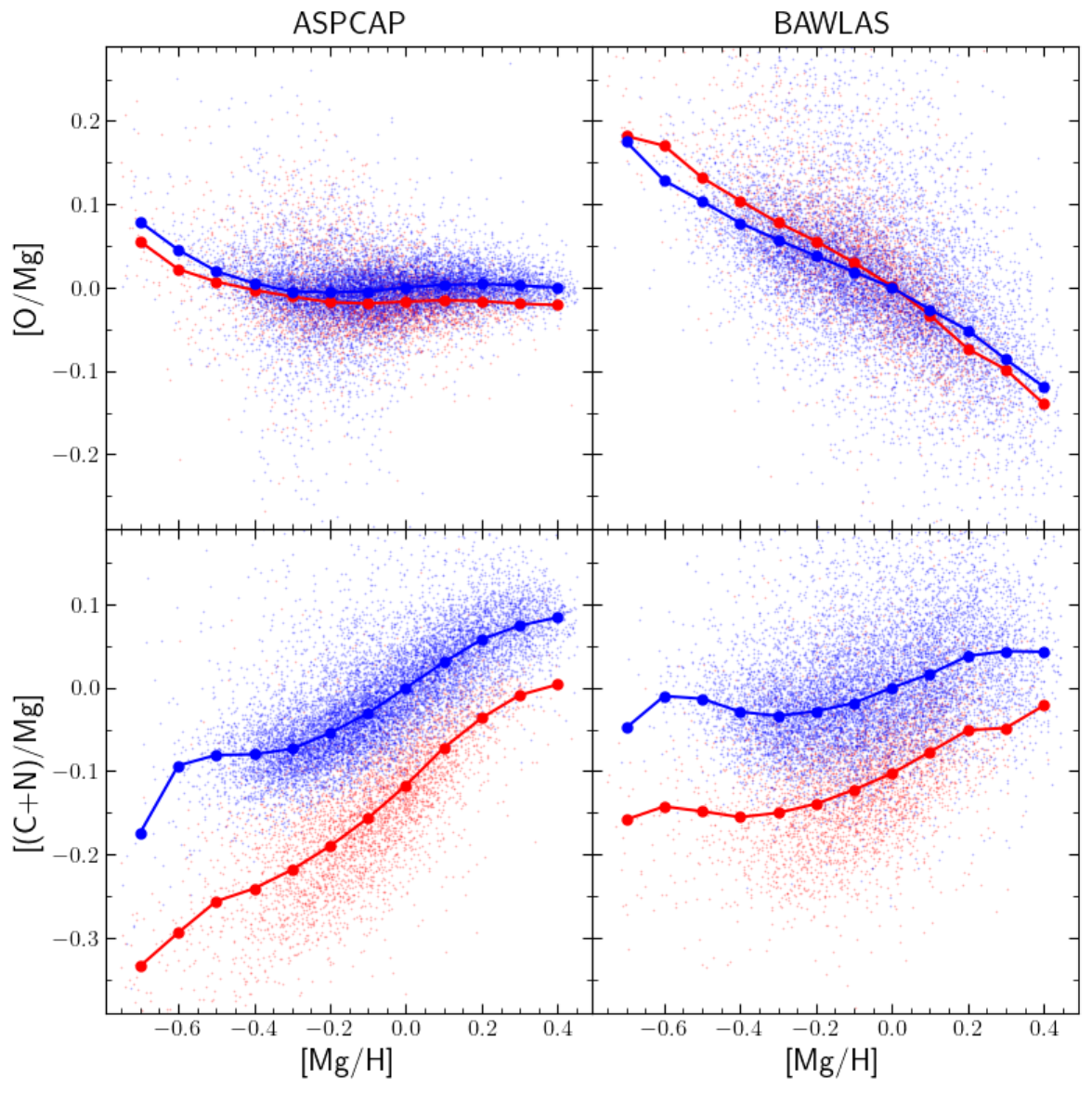}
    \caption{[X/Mg] vs. [Mg/H] for ASPCAP (left column) and BAWLAS (right column) measurements of O (top row) and C+N (bottom row). Stars are randomly downsampled by 10 for ASPCAP measurements and 5 for BAWLAS measurements to reduce crowding. Colors and symbols are identical to that of Figure \ref{fig:CuMg}. This figure uses data from the calibration sample.}
    \label{fig:ASPCAPvsBAWLAS}
\end{figure*}

With both $\log(g)$ and zero-point corrections in hand, we calculate the median low- and high-Ia trends in 0.1 dex [Mg/H] bins from [X/H]$_{\text{corr}}$ (Equation \ref{eq:xh_corr}) for the SNR $>$ 100 main survey disk sample ($N=151,564$, Section \ref{subsec: sample}). We plot the median trends for the new BAWLAS-only element Cu in Figure \ref{fig:CuMg}. The trend with increasing metallicity for [Mg/H] $>-0.2$ and the $\approx$0.2 dex separation between the high- and low-Ia tracks are similar to that of Co and Ni, the other Fe-peak elements heavier than Fe, in this metallicity range, but the Cu trend is steeper. This suggests that Cu has a similar level of delayed contributions as Co and Ni, but potentially a stronger metallicity dependence at higher metallicities. In this [Mg/H] $>-0.2$ metallicity range, our results also qualitatively match optical results from GALAH. At [Mg/H] $<-0.2$, the [Cu/Mg] abundance increases with decreasing [Mg/H], a trend only shared with V (in both the ASPCAP and BAWLAS measurements of V). However, both elements have weak lines in the APOGEE wavelengths, and the BAWLAS median Cu trend does not match those from optical GALAH spectra at the same [Mg/H] \citep[e.g.,][]{Griffith2022}, so the observed decreasing trend may be somewhat biased towards stars with higher Cu abundances and stronger lines at lower metallicities. \citet{Hayes2022} compared the overall BAWLAS Cu distribution to literature measurements from high-resolution spectra and found similar trends, although Fe was used as the reference element. The median trends of the other elements for both ASPCAP and BAWLAS are similar to \citetalias{Weinberg2022} so we do not reproduce them here, except for C+N and O (Figure \ref{fig:ASPCAPvsBAWLAS}) which we discuss further in Section \ref{subsec:resid_abunds}.

We then use these medians to derive the process vectors $q^{X}_{\text{cc}}(z)$ and $q^{X}_{\text{Ia}}(z)$ for all elements following Equations 25 and 26 from \citetalias{Weinberg2022}. We report the values of $q^{X}_{\text{cc}}(z)$ and $q^{X}_{\text{Ia}}(z)$ in Tables \ref{table:qcc} and \ref{table:qIa} in Appendix \ref{appx:q_vals}.

\citet{Griffith2023} propose an alternative method of inferring process vectors that does not involve median sequences. In the disk regime where these sequences are clearly defined, \citet{Griffith2023} show that their method gives process vectors nearly identical to those of the \citetalias{Weinberg2022} method, but their approach adapts more easily to small samples or to the low metallicity regime where [$\alpha$/Fe] bimodality is not obvious. Since our calibration sample is limited to the MW disk, we use the \citetalias{Weinberg2022} method in this work, but we expect that applying the \citet{Griffith2023} method to our $\log(g)$-calibrated abundances should yield similar process vectors.

\subsection{Two-process Amplitudes, $A_{\textrm{cc}}$ and $A_{\textrm{Ia}}$}\label{subsec:2proc_amplitudes}
Using Equation \ref{eq:2proc} and our derived $q^{X}_{\text{cc}}(z)$ and $q^{X}_{\text{Ia}}(z)$ vectors, we can now fit for the process amplitudes $\Acc$ and $\AIa$ for all 310,427 observations of 288,789 unique stars in our final sample. First, $C^X_{\log(g)}$ is calculated for each element of each star, and added to the raw reported abundance along with the appropriate $C^X_{\text{ZP}}$ for that element. Following \citetalias{Weinberg2022}, we infer $\Acc$ and $\AIa$ from six APOGEE elements (Mg, O, Si, Ca, Fe, Ni) that have small observational errors, a range of relative contributions from CCSN/SNIa, and are theoretically unlikely to have contributions from non-CCSN/SNIa sources. Specifically, we use Equation \ref{eq:2proc} to calculate a predicted two-process [X/H] for X = Mg, O, Si, Ca, Fe, Ni, then use the observed, corrected [X/H] (Equation \ref{eq:xh_corr}) and reported measurement errors to calculate $\chi^2$. For each star, we minimize the $\chi^2$ value of the fit to these six elements using an iterative process. Starting with an initial guess of $\Acc$ and $\AIa$ based only on [Mg/H], [Fe/Mg], and the plateau $\text{[Fe/Mg]}_{\text{pl}} = -\text{[Mg/Fe]}_{\text{pl}} = -0.3$ (\citetalias{Weinberg2022} Eqs. 13, 18), we hold either $\Acc$ or $\AIa$ fixed while minimizing the other. To avoid outliers impacting the best-fit $\Acc$ and $\AIa$ values, O, Si, Ca, and/or Ni were eliminated from the fitting process if their abundance deviated by $>$5$\sigma$ from the initial guess. This six-element method reduces the effect of measurement aberrations in the residual abundances compared to using only Fe and Mg \citep[e.g.,][]{Griffith2019,Weinberg2019}, as demonstrated in \citetalias{Weinberg2022}.

\begin{figure*}[!th]
    \centering
    \includegraphics[width=\linewidth]{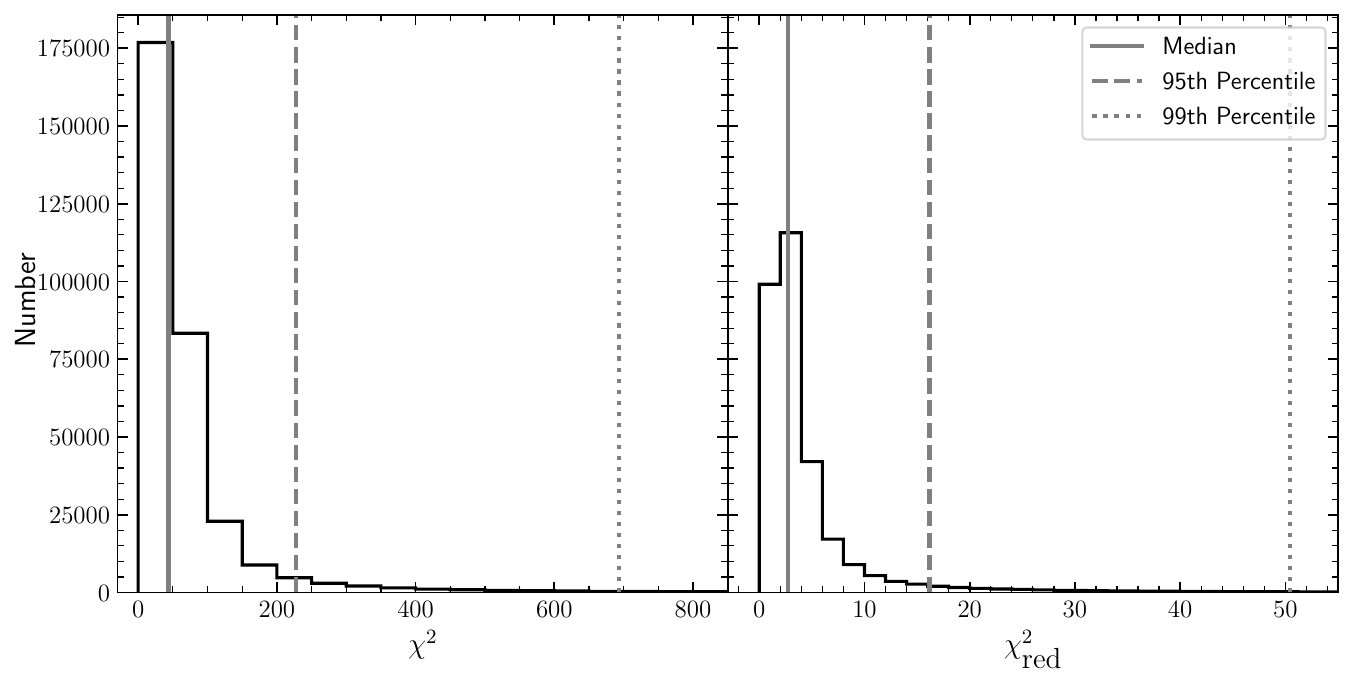}
    \caption{Histograms of the $\chi^2$ (left) and reduced $\chi^2$ ($\chi^2_{\text{red}}$, right) of the two-process fits to our final sample ($N=\text{310,427}$). $\chi^2_{\text{red}}$ is defined by Equation \ref{eq:red_chi2}. The solid, dashed, and dotted gray lines mark the median, 95th percentile, and 99th percentile.}
    \label{fig:chi2}
\end{figure*}

Once the best-fit process amplitudes are derived for each star, we use these amplitudes and the global process vectors $q^{X}_{\text{cc}}(z)$ and $q^{X}_{\text{Ia}}(z)$ in Equation \ref{eq:2proc} to predict the abundances for \textit{all} elements. The $\chi^2$ value reported in our catalog for each star is calculated from all valid (non-NaN, i.e., existing and unflagged) abundances. We also report a reduced $\chi^2$ for each star in our catalog, defined as
\begin{equation}\label{eq:red_chi2}
    \chi^2_{\text{red}} = \frac{\chi^2}{N_{\text{elem}}-2}
\end{equation}
where $N_{\text{elem}}$ is the number of elements with valid abundance measurements for that star. The model has two free parameters, $\Acc$ and $\AIa$, so we subtract 2 to obtain the total degrees of freedom.

The distributions of $\chi^2$ and $\chi^2_{\text{red}}$ for the final sample are shown in Figure \ref{fig:chi2}. The median $\chi^2$ value is $\approx$44 for a maximum of 21 degrees of freedom (23 total elements, as we considered ASPCAP and BAWLAS measurements separately, with two free parameters), and the median $\chi^2_{\text{red}}$ is $\approx$2.7. This suggests that the true distribution of abundances has intrinsic scatter not captured by the two-process model and/or that the true abundance uncertainties are larger than those predicted for Gaussian noise with the reported measurement error. \citetalias{Weinberg2022} reported a median $\chi^2$ of 30 for 14 degrees of freedom, which is a similar ratio to our results. Based on correlations of residual abundances, \citet{Ting2022}, \citet{Griffith2022}, and \citetalias{Weinberg2022} demonstrate that at least some of the excess $\chi^2$ is a consequence of physical intrinsic scatter rather than underestimated statistical errors.

\begin{figure}[!th]
    \centering
    \includegraphics[width=\linewidth]{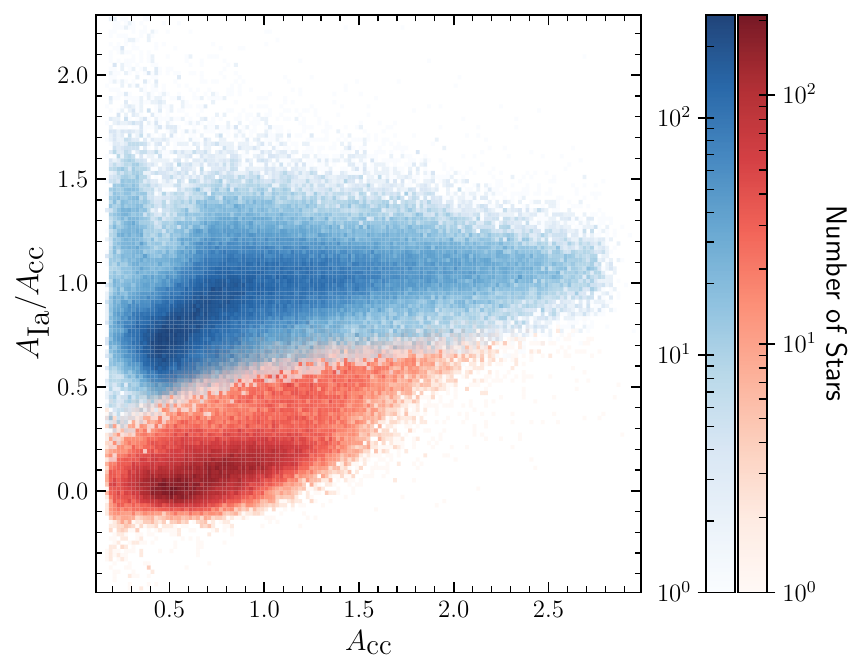}
    \caption{The process amplitudes of the full sample ($N=\text{310,427}$), plotted as $\AIa/\Acc$ vs. $\Acc$. The high-Ia stars are in blue and the low-Ia stars in red. The bin size is 0.02 along both axes.}
    \label{fig:aIa_vs_aCC}
\end{figure}

In Figure \ref{fig:aIa_vs_aCC}, we plot the distribution of best-fit $\AIa/\Acc$ vs. $\Acc$ values. The high-Ia (in blue) and low-Ia (in red) populations separate, similar to a typical [Mg/Fe] vs. [Fe/H] plot but more closely analogous to [Fe/Mg] vs. [Mg/H], with $\Acc$ providing a measure of metallicity as traced by CCSN elements. The division of the two populations at $\AIa/\Acc \approx 0.5$ reflects the definition of the populations (Equation \ref{eq:alpha_split}). Unlike the flat plateau at $\text{[Fe/Mg]}=-0.3$ for the low-Ia population in an [Fe/Mg] vs. [Mg/H] plot, the low-Ia population shows a steady rise in $\AIa/\Acc$ as $\Acc$ increases. The high-Ia population shows a similar but steeper increase in $\AIa/\Acc$ before plateauing at $\AIa/\Acc\approx1$. The $\AIa/\Acc$ rise in both populations traces an increase in time-delayed SNIa enrichment over time, and the shallower rise in the high-Ia trend plausibly results from higher star formation efficiency, which allows the population to reach higher $\Acc$ within the characteristic SNIa delay time $\tau_{\text{Ia}}\approx1-2$ Gyr. The flat plateau at $\AIa/\Acc\approx1$ suggests that these stars form when the CCSN and SNIa rates are approximately equal to each other, producing near-solar abundance ratios. These general trends are more obvious in the calibration sample, which consists only of MW disk stars. The high $\AIa/\Acc$ feature at $\Acc<0.5$ is not seen in the disk and consists primarily of dwarf galaxy members (Section \ref{subsec:gals}). The detailed appearance of this diagram depends on our key model assumption that [Mg/Fe] = 0.3 is the ratio produced by pure CCSN enrichment ($\AIa=0$) (Section \ref{subsec:2proc_vectors}).

\subsection{Residual Abundances, $\Delta$[X/H]}\label{subsec:resid_abunds}
With the two-process vectors and fitted $\Acc$ and $\AIa$, we can predict all of a star's ASPCAP and BAWLAS abundances, but we do not expect these predictions to be perfect. Some deviations will arise from problematic spectra and/or measurement errors. However, other deviations, particularly those common to certain populations of stars, may be real and encode information about additional nucleosynthetic sources, chemically peculiar stars, and abundance trends in stellar populations.

\begin{figure*}[!th]
    \centering
    \includegraphics[width=0.9\linewidth]{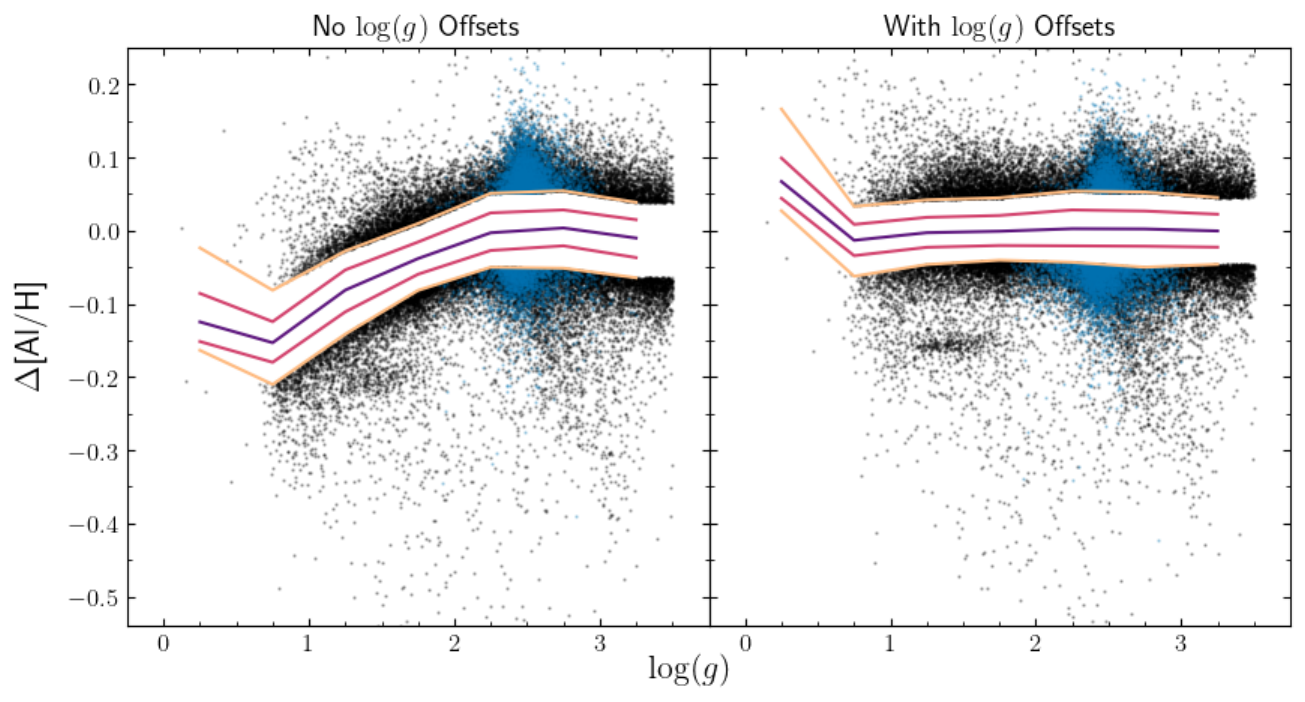}
    \caption{Residual abundances (Section \ref{subsec:resid_abunds}) of Al as a function of $\log(g)$ before (left) and after (right) applying the $\log(g)$ calibration offsets $C_{\log(g)}$. The calibration offsets were applied to abundances prior to calculation of the median [Al/Mg] vs. [Mg/H] tracks, so that the left and right plots use different sets of two-process vectors. The purple, magenta, and peach lines indicate the median, 25th/75th percentiles, and 10th/90th percentiles of each $\log(g)$ bin. Stars falling outside the 10th/90th percentiles are plotted as points; red clump stars are indicated in blue. For the $\log(g)$ bins centered at $\log(g)\geq0.75$, our calibrations clearly flatten out the overall trend with $\log(g)$ around median $\Delta\text{[Al/H]}=0$. This figure uses the calibration sample (Section \ref{subsec: sample}).
    }
    \label{fig:resids_beforeafter_logg}
\end{figure*}

Following the convention of \citetalias{Weinberg2022} and \citet{Griffith2022}, we define residual abundances in the rest of this work as observed$-$predicted, specifically:
\begin{equation} \label{eq:residual_def}
    \Delta\text{[X/H]} = \text{[X/H]}_{\text{corr}} - \text{[X/H]}_{\text{2proc}},
\end{equation}
where [X/H]$_{\text{corr}}$ is defined in Equation \ref{eq:xh_corr} and [X/H]$_{\text{2proc}}$ is the abundance predicted by the two-process model (Equation \ref{eq:2proc}). The catalog of $\Acc$, $\AIa$, and residual abundances for our sample of 288,789 stars is a major product of this paper. We describe how to access and use it in Appendix \ref{appx:catalog}.

In Figure \ref{fig:resids_beforeafter_logg}, we plot the residual abundances of Al, $\Delta$[Al/H], with and without applying the $\log(g)$ calibrations offsets $C_{\log(g)}$ (Section \ref{sec:logg}). The figure illustrates that our calibrations are effective at removing overall trends with $\log(g)$ in the residual abundances, as expected. We have also confirmed that they remove trends with $\Teff$. No $\log(g)=0.25$ bin at any metallicity contained enough stars in the calibration sample to calculate $\log(g)$ offsets, resulting in a small number of stars with $\log(g)=0-0.5$ using offsets from the $\log(g)=0.75$ bin at matching [Mg/H].

\begin{figure*}[!ht]
    \centering
    \includegraphics[width=\linewidth]{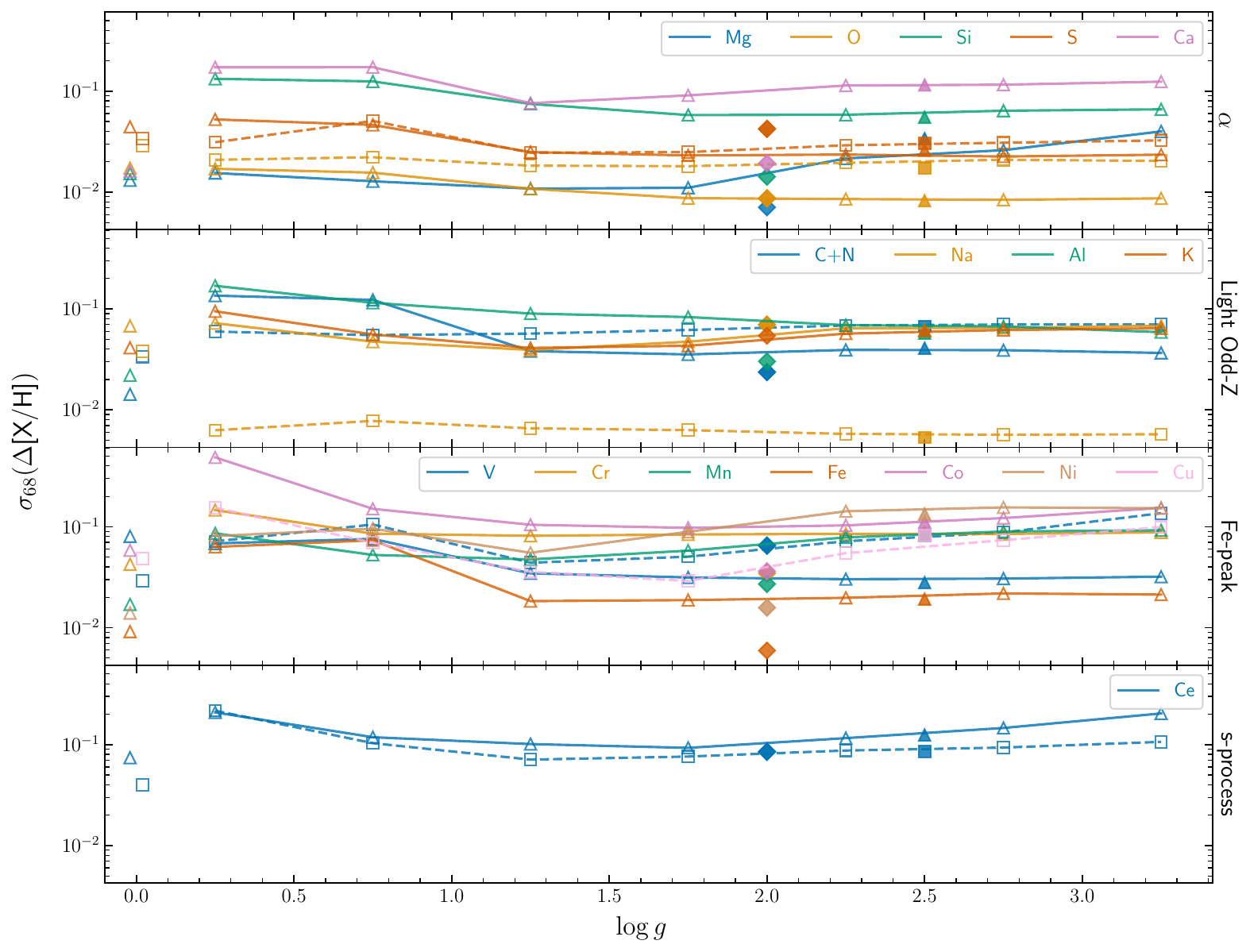}
    \caption{Scatter in two-process residuals as a function of $\log(g)$, quantified as half of the difference between 84th and 16th percentiles ($\pm1\sigma$ of a Gaussian distribution) to reduce the effect of outliers. ASPCAP abundances are shown in solid lines and triangles. BAWLAS abundances are shown in dashed lines and squares. Filled triangles/squares indicate the red clump sample, which has median $\log(g) \approx 2.5$. Filled diamonds at $\log(g)=2.0$ show $\sigma_{68}$ of the two-process residuals from the smaller sample in \citetalias{Weinberg2022}, with $\log(g) = 1-2.5$, for comparison. The points marked around $\log(g)=0$ represent the median measurement error for each element; the colors and symbols of these points match those of the two-process residual scatter. Data in this figure are from the full sample.}
    \label{fig:2proc_residual_scatter}
\end{figure*}

In Figure \ref{fig:2proc_residual_scatter}, we plot the dispersion of two-process residuals for each element as a function of $\log(g)$. To eliminate sensitivity to outliers, we compute the dispersion $\sigma_{68}$ as half the difference between the 84$^{\text{th}}$ and 16$^{\text{th}}$ percentile values of $\Delta$[X/H] in a given $\log(g)$ bin. The scatter in the residuals is fairly flat over our sample's $\log(g)$ range overall, although there is a slight increase in the lowest $\log(g)$ bins, which have the fewest stars. Compared to \citetalias{Weinberg2022}, who used a more limited range in surface gravity and effective temperature ($\log(g) = 1-2.5$ and $\Teff = 4000-4600$) and thus had a sample $\approx$8 times smaller than in this work, the dispersion in our sample is only slightly higher on average. If we did not introduce $\log(g)$ calibration offsets, the residual abundance scatter would be substantially larger because it would be inflated by systematic errors over the large $\log(g)$ range of our sample. This is best illustrated by six ASPCAP elements that show a $>$25\% decrease in $\sigma_{68}$ after applying the $\log(g)$ calibration offsets: O (0.007 dex, 26.4\%), Al (0.014 dex, 27.3\%), Ca (0.010 dex, 28.7\%), V (0.055 dex, 31.6\%), and Mn (0.020 dex, 40.5\%). The flatness of the scatter trends in Figure \ref{fig:2proc_residual_scatter} and the similarity to the scatter found in the narrower sample of \citetalias{Weinberg2022} demonstrate the effectiveness of the calibration offsets derived from the median abundance trends. We also see generally higher $\sigma_{68}$ at low [Mg/H], with the most notable increases at [Mg/H] $<-0.4$. This effect is most likely due to lower measurement precision at lower metallicities, but we did not distinguish measurement uncertainty vs. intrinsic scatter in our calculation of $\sigma_{68}$.

For elements shared between ASPCAP and BAWLAS, the BAWLAS measurements generally show a similar amount of scatter. The dispersion in V shows the largest reduction from ASPCAP to BAWLAS, especially at higher $\log(g)$, suggesting that a significant amount of the total dispersion in V is observational rather than intrinsic. On the other hand, C+N and O show notably higher two-process residual scatter in BAWLAS than ASPCAP. This increased dispersion can be seen when comparing [X/Mg] values around the median for these elements in Figure \ref{fig:ASPCAPvsBAWLAS}, and BAWLAS reports higher median measurement errors than ASPCAP for both elements.

Furthermore, the top row in Figure \ref{fig:ASPCAPvsBAWLAS} shows that O has a dramatically different median trend with [Mg/H] in BAWLAS than ASPCAP: the ASPCAP median trend in [O/Mg] is nearly flat for $\text{[Mg/H]}\gtrsim-0.4$ while BAWLAS [O/Mg] steeply declines with increasing [Mg/H]. This difference has long been observed between O abundances measured from near-IR and optical spectra \citep[e.g.,][]{Bensby2014,Griffith2022}, but BAWLAS abundances are measured from the same near-IR APOGEE spectra as ASPCAP. We note that the optical O measurements use atomic lines, while the near-IR ASPCAP and BAWLAS measurements use molecular OH lines. \citet{Hayes2022} attribute the difference between the ASPCAP and BAWLAS O trends to the BAWLAS analysis using calibrated stellar parameters for $\Teff$ and $\log(g)$, whereas ASPCAP used the uncalibrated parameters. In essence, ASPCAP adopts the $\Teff$ and $\log(g)$ that gives the best fit to the whole APOGEE spectrum simultaneously with several other stellar parameters (including overall metallicity [M/H] and overall $\alpha$-element abundance [$\alpha$/H], which includes O; see \citet{GarcíaPérez2016} for details). BAWLAS adopts adjusted stellar parameters that have been calibrated using external references (e.g., astroseismic surface gravities, photometric temperatures). It is not obvious which approach is preferable for deriving abundances, or why systematics associated with the use of calibrated vs. uncalibrated parameters would reproduce the optical abundance trends.

The closeness of the high- and low-Ia sequences indicates that O production is dominated by prompt CCSN for both ASPCAP and BAWLAS determinations, but ASPCAP's mostly-flat slope implies metallicity-independent yields while the downward slope of BAWLAS implies that O yields decrease with increasing metallicity. The crossing of the high-Ia sequence above the low-Ia sequence in BAWLAS O at $\text{[Mg/H]}=0$ leads to negative $q_{\text{Ia}}$ values at sub-solar metallicities. Negative $q_{\text{Ia}}$ values are technically unphysical, according to the two-process model assumption that $q^X_{\text{Ia}}=0$ represents pure CCSN enrichment for element $X$, but the high- and low-Ia sequences are close enough to one another that the negative values are most likely the result of scatter in the BAWLAS O measurements. A smaller effect seems to be present for C+N, where the median [(C+N)/Mg] trend more steeply increases with [Mg/H] in ASPCAP than BAWLAS.

For some elements the scatter of the two-process residual is similar in magnitude to the typical reported measurement error. However, some elements show scatter significantly above the measurement errors (e.g., Ca, Na, Mn), suggesting a substantial contribution of intrinsic scatter. The estimation of intrinsic scatter and correlation of residuals is discussed at length by \citetalias{Weinberg2022} (see their Figures 13 and 15), so we will not repeat this analysis here. Other discussion of the intrinsic scatter of APOGEE abundances around a 2-parameter fit can be found in \citet{Ting2022} and \citet{Ness2022}.

\section{Residual Abundances of Stellar Populations} \label{sec:populations}

One application of the two-process model and residual abundances is to examine differences in the enrichment patterns of different stellar populations, which may be defined by spatial, kinematic, and age cuts or membership in star clusters or satellite galaxies. These populations often span a range in overall metallicity and [$\alpha$/Fe] ratio, producing correlated variations across many elements, so a traditional ``chemical tagging" approach risks counting the same physical effect many times in different elements. Two-process modeling encapsulates these correlated variations in the $\Acc$ and $\AIa$ parameters, and residual abundances can reveal subtle deviations that could reflect different levels of AGB enrichment, differences in supernova populations, or other effects. Because we ``train" the two-process vector on the disk median sequences, residuals of stellar populations reveal systematic differences from typical disk stars of matched [Mg/H] and [$\alpha$/Fe]. \citetalias{Weinberg2022} applied this approach to find 0.05-0.2 dex differences for multiple elements in stellar populations from the Large Magellanic Cloud (LMC) and the Gaia-Sausage/Enceladus (GSE) remnant, and much larger deviations for $\omega$ Cen. Conversely, they found no differences (at the 0.01-0.02 dex level) between stars at $R = 15-17$ kpc and stars in the main body of the disk.

Our stellar sample is $\approx$8 times larger than that of \citetalias{Weinberg2022} overall, and the gain is even larger for star clusters and dwarf satellites, where many of the APOGEE targets lie outside the $\log(g)$ and $\Teff$ cuts used by \citetalias{Weinberg2022}. As a first application of our expanded catalog, therefore, we investigate residual abundances of a selection of open clusters, globular clusters, and MW satellites. To ensure robust residual abundance measurements, we apply a S/N$>$100 cut for the following sections, but utilizing the lower S/N$>$80 limit of the overall catalog does not meaningfully change any results.

\subsection{Open Clusters} \label{subsec:OCs}

Open clusters are groups of stars that formed from the same molecular cloud at the same time in Galactic history, and are thus expected to have nearly uniform chemical abundances. Previous work has shown that members of open clusters generally show chemical homogeneity within 0.02-0.04 dex \citep[e.g.,][]{DeSilva2006,Liu2016,Bovy2016,Casamiquela2020,Poovelil2020,Ness2022}. Within the context of the two-process model, we would expect that any enhancements or depletions seen in the residual abundances will be shared by the member stars of an open cluster. Dating of open clusters via color-magnitude diagrams is likely to be more accurate than isochrone dating of individual stars. Combining open cluster ages with the use of median deviations to reduce the impact of potential measurement errors and statistical uncertainties within a chemically homogeneous population can provide interesting and robust insights into changes in disk abundance patterns with age.

\citet{Griffith2022} analyzed the residual abundances of 14 open clusters using GALAH. They found that younger clusters have larger residual abundances than older clusters as a general trend and noted significant deviations in the elements O, Ca, K, Cu, Y, and Ba. The positive Y and Ba residuals were consistent with previous literature, including super-solar abundances of these elements in \citet{Spina2021} and enhancement of Y and Ba in young open clusters \citep[e.g.,][]{Baratella2021,Casamiquela2021}, which suggest an enhanced \textit{s}-process \citep{DOrazi2009}. O, Ca, and K had positive residuals (i.e. enhancement compared to field stars) that were stronger in the younger clusters, while Cu had the opposite trend, showing more negative residuals (i.e. depletion compared to field stars) in the younger clusters. Based on these results, \citet{Griffith2022} argue that open clusters' unique enrichment patterns and age-correlated residuals make them promising candidates for testing the feasibility of chemical tagging.

Here, we perform a similar analysis using our APOGEE sample. The differences in surveys used and sample cuts result in several important differences between our analysis and \citet{Griffith2022}:
\begin{enumerate}
    \item GALAH is an optical survey while APOGEE is a near-IR survey, so the lines available in each spectral range are different, and the resulting analyses from the different surveys will result in different systematics. This is well-illustrated by the notable negative trend in [O/Mg] vs. [Mg/H] in GALAH measurements not present in ASPCAP abundances. Interestingly, this trend is also seen in the BAWLAS measurements of APOGEE spectra. In theory, the \textit{true} abundances of elements shared by both surveys should be the same. 
    \item \citet{Griffith2022} selected dwarf and subgiant stars with $\log(g) = 3.5-4.5$ and $\Teff = 4200-6200$ K. In contrast, we select evolved stars with $\log(g) = 0-3.5$ and $\Teff = 3000-5000$ K.
    \item The \citet{Griffith2022} open cluster sample contains many more young $<$1 Gyr clusters. The youngest cluster in our sample, NGC 6705, would be the sixth oldest cluster in the \citet{Griffith2022} sample. We also have two clusters older than the oldest cluster in \citet{Griffith2022}. The difference in age distributions may reflect the evolutionary state of the stars in our respective samples.
\end{enumerate}

\begin{figure*}[!h]
    \centering
    \includegraphics[width=\linewidth]{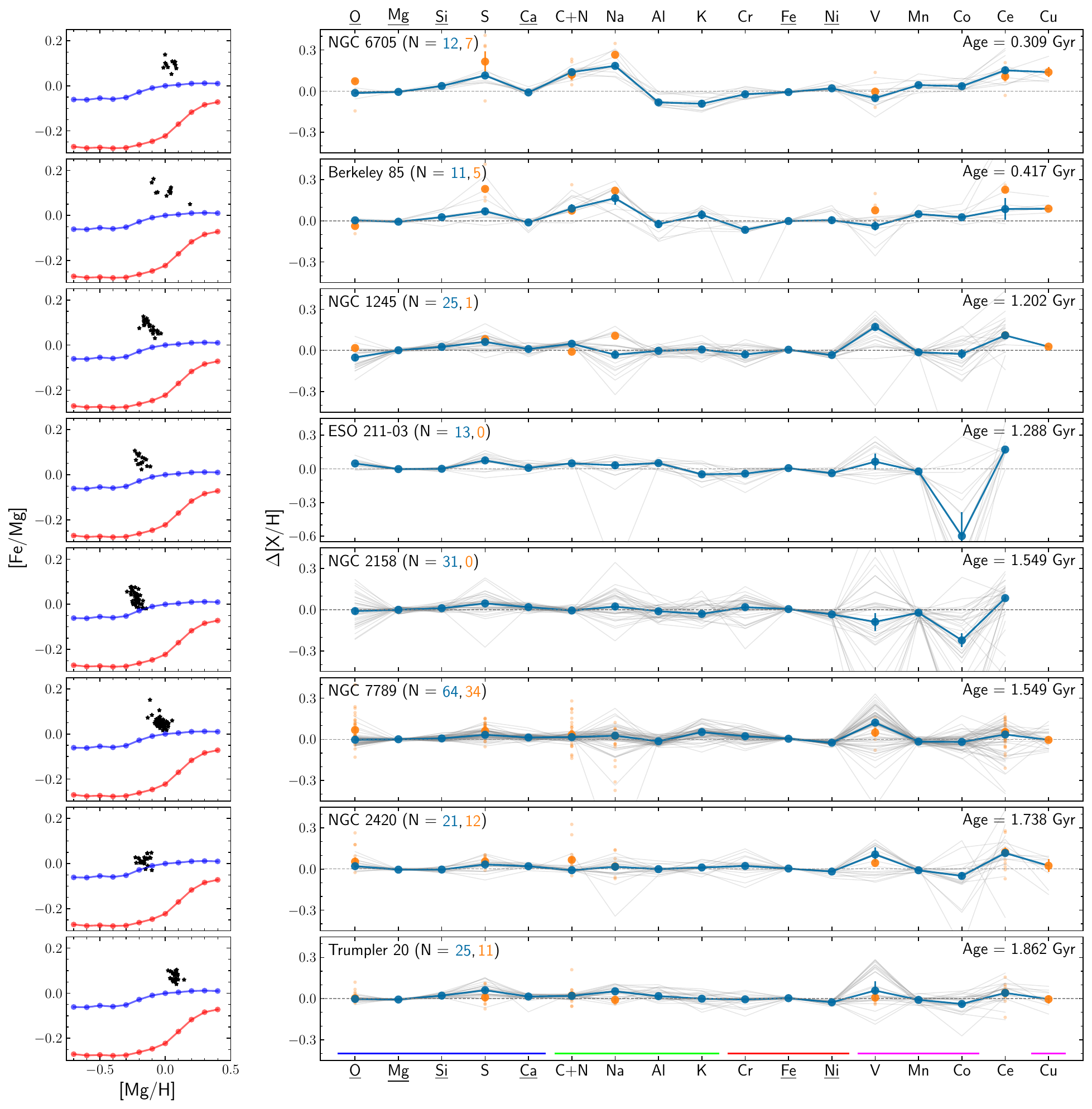}
    \caption{\textit{Left}: [Fe/Mg] vs. [Mg/H] abundances for the 8 youngest ($\leq$2 Gyr) open clusters in our sample. The median low-Ia (red) and high-Ia (blue) sequences for the calibration (disk) sample are shown for reference. Individual stars in each cluster are shown by black stars. \textit{Right}: residual abundances of the 8 youngest open clusters in our sample. Light gray lines show the residuals for each individual member star for the ASPCAP elements and BAWLAS Cu. Individual residuals for BAWLAS O, S, C+N, Na, V, and Ce are shown in small orange points. Large blue points indicate the median deviation of ASPCAP measurements for each element and the corresponding ASPCAP sample size is shown in blue text. Large orange points indicate the median of BAWLAS measurements and the BAWLAS sample size is shown in orange text. We note that the sample size for a specific element may be less than the listed number if any individual star's measurement was flagged (Section \ref{sec:data}). A blue line has been plotted connecting different elements' median residuals, with preference to the ASPCAP measurement due to larger sample size, to guide the eye. Error bars show the standard deviation on the median computed from 1000 bootstrap resamplings of each cluster. When not visible, these error bars are smaller than the plotted points. Colored lines along the bottom of the lower panel group $\alpha$ elements (blue), light odd-Z elements (green), even-Z iron-peak elements (red), and odd-Z iron-peak elements (magenta). Note that the panel for ESO 211-03 has a different vertical range.}
    \label{fig:2proc_OCs_1}
\end{figure*}

To obtain our open cluster sample, we cross-match our two-process residual catalog with the Open Cluster Chemical Abundances and Mapping (OCCAM) survey for DR17 \citep{Myers2022}. We consider only clusters with at least 8 member stars in OCCAM and designated as ``high-quality" based on their color-magnitude diagrams (Table 1, \citealt{Myers2022}). We also only consider stars in those clusters within 3$\sigma$ of the cluster mean in proper motion, radial velocity, and [Fe/H] (following the recommendations in \citealt{Myers2022}). We find 26 open clusters in OCCAM represented in our sample with these constraints, and ultimately select 14 of the 15 clusters that have $N>10$ stars with two-process residuals. Dolidze 41 was excluded because it contained exactly the same stars as Berkeley 85. The ages of the clusters are also taken from the OCCAM catalog. The sample of stars within each OC generally spans a reasonably wide range of $\Teff$-$\log(g)$ space, with the exception of NGC 1245 and ESO 211-03, whose stars are both concentrated around $\Teff\approx5100-5200$ K and $\log(g)\approx2.75-3$. However, these two clusters show different residual abundance patterns from each other (Figure \ref{fig:2proc_OCs_1}), so there do not appear to be remaining systematics from $\log(g)$ or $\Teff$.

\begin{figure*}[!ht]
    \centering
    \includegraphics[width=\linewidth]{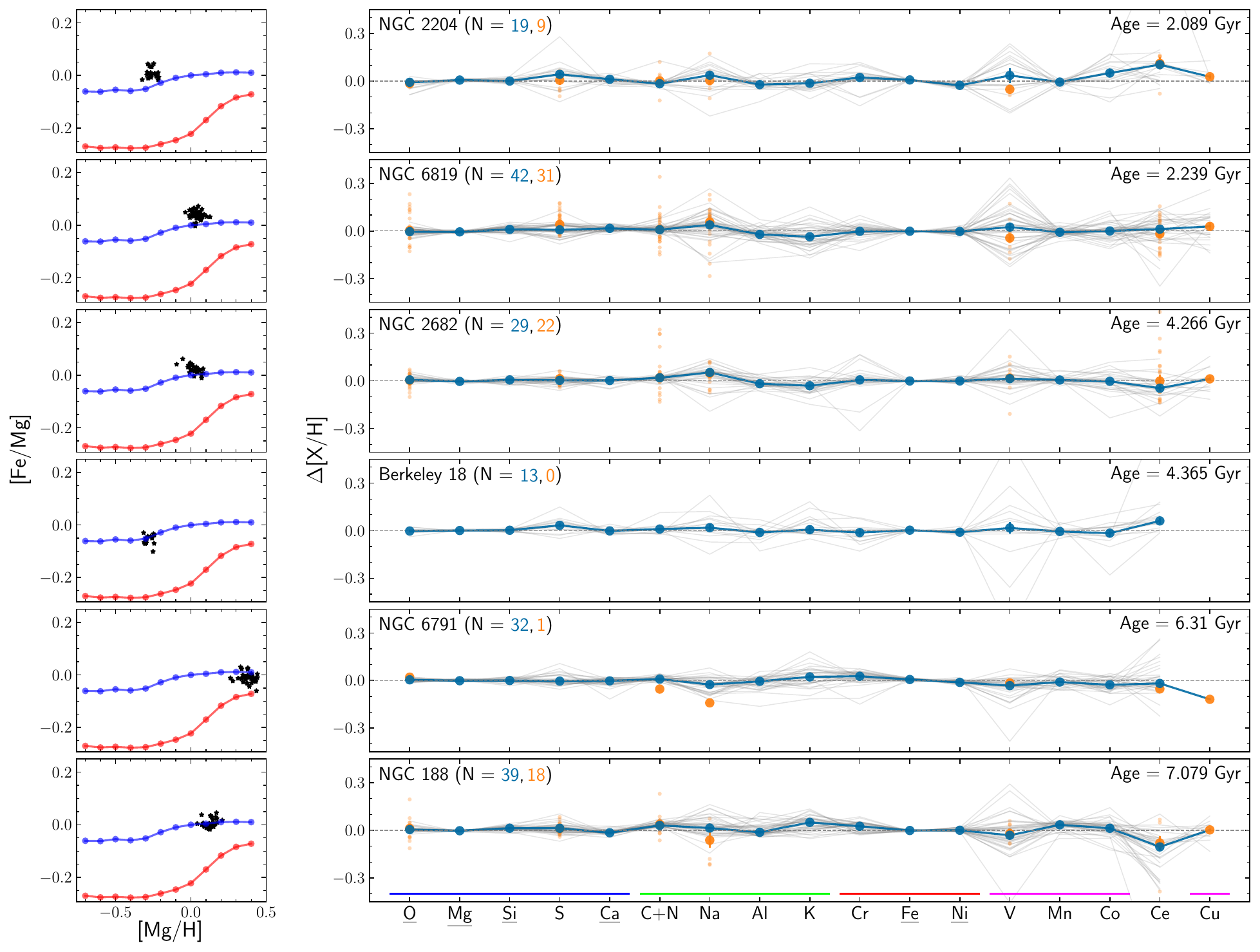}
    \caption{Same as Figure \ref{fig:2proc_OCs_2} for the 6 oldest ($\geq$2 Gyr) open clusters in our sample.}
    \label{fig:2proc_OCs_2}
\end{figure*}

In the left panels of Figures \ref{fig:2proc_OCs_1} and \ref{fig:2proc_OCs_2}, we plot [Fe/Mg] vs. [Mg/H] for all stars in each of the 14 clusters in our sample. We notice immediately that the stars in all of the clusters lie on or above the high-Ia median sequence, with the older clusters having lower [Fe/Mg] values and lying closer to the median sequence than the younger clusters. This generally makes sense because the gas from which the youngest clusters formed would likely have had more time to be enriched by delayed SNIa contributions. The member stars in each cluster are generally well-concentrated in [Fe/Mg]-[Mg/H] space as expected. However, the observed spread within each cluster is $\approx$0.1-0.2 dex, larger than the inhomogeneity reported in higher SNR studies. Furthermore, the measurement errors on ASPCAP [Fe/H] and [Mg/Fe], when added in quadrature, yield median [Mg/H] uncertainties of $\approx$0.012-0.020 dex across our open cluster sample and are insufficient to fully explain the observed spread.

\begin{figure*}[!ht]
    \centering
    \includegraphics[width=0.8\linewidth]{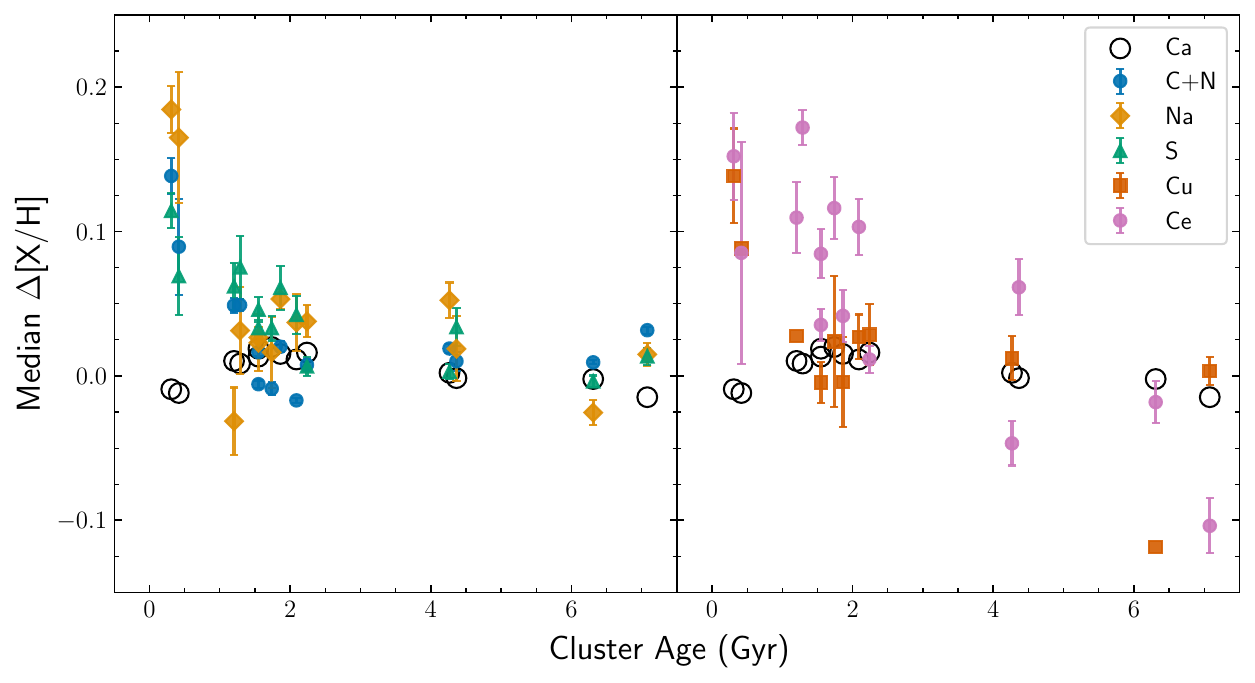}
    \caption{Residual abundances of each open cluster in our sample as a function of cluster age for selected elements. The division into two panels is only to improve visibility. Blue circles, yellow diamonds, green triangles, orange squares, and pink circles represent the median residual abundance of C+N, Na, S, Cu, and Ce respectively. The error bars show the standard deviation on the median computed from 1000 bootstrap resamplings of each cluster. ASPCAP measurements are used for all elements except the BAWLAS-only element Cu. The median residual abundance of Ca is also shown in open black circles on both plots as an example of an element with little residual abundance variation with age in our sample, and the bootstrapped uncertainties are smaller than the marker size.}
    \label{fig:OC_resids_vs_age}
\end{figure*}

The right panels of Figures \ref{fig:2proc_OCs_1} and \ref{fig:2proc_OCs_2} show the residual abundances of our 14 open clusters. We include error bars on median residuals based on bootstrap resampling, but these are usually smaller than the plotted points. This plot demonstrates the value of investigating median residuals, as differences at the 0.1 dex level can be detected at high significance even when individual stars may have 0.05-dex level observational uncertainties (which one can judge from the grey lines showing individual stars). We generally find a similar trend to \citet{Griffith2022}, where the youngest clusters show the strongest residuals while the older clusters show much smaller residuals, if any. In our sample, the abundances become close to that of field stars by $\approx$1.7-1.8 Gyr. The two youngest clusters in our sample, NGC 6705 and Berkeley 85, show enhancements in S, C+N, Na, Cu, and Ce. We plot these elements as a function of cluster age in Figure \ref{fig:OC_resids_vs_age}. 

We observe Ce enhancement of 0.05-0.2 dex in all clusters with ages up to $\approx$2 Gyr (all clusters in Figure \ref{fig:2proc_OCs_1}, and NGC 2204), and in Berkeley 18 (age = 4.3 Gyr). This age trend is different from other elements shown in Figure \ref{fig:OC_resids_vs_age}, which only show clear enhancement in the two youngest clusters. The enhancement in Ce is consistent with \citet{SalesSilva2022}, who found higher [Ce/Fe] and [Ce/$\alpha$] in open clusters younger than 4 Gyr, as well as enhancement in other \textit{s}-process elements like Y and Ba in young open clusters \citep[e.g.,][]{Baratella2021,Casamiquela2021}. Furthermore, the dispersion in Ce enhancement in the $\approx$2 Gyr old OCs is resembles a similar trend in Ba observed by \citet{Mishenina2015}, which they argue is a signature of an intermediate neutron-capture process (``\textit{i}-process"). We would likely need to compare our Ce residuals to other neutron-capture elements to confirm or rebut an \textit{i}-process nucleosynthesis signature. 

Ce enhancement sometimes arises in concert with C+N enhancement as a consequence of binary mass transfer (e.g., \citealt{Smith2000}, and discussed in the context of residual abundances in \citetalias{Weinberg2022}). However, we expect this phenomenon to affect only a small fraction of stars, so the open cluster Ce enhancement is likely present in birth abundances. Outside clusters, \citet{Casali2023} studied low-$\alpha$ stars with astroseismic ages and found that younger stars have higher [Ce/Fe] at fixed [Fe/H] and higher [Ce/$\alpha$] at fixed radial position. In residual abundances, \citetalias{Weinberg2022} find a trend of enhanced Ce in young APOGEE field stars, and \cite{Griffith2022} find a similar trend for Y and Ba in GALAH, so a plausible interpretation is that both cluster and field stars reflect a general increase of AGB enrichment in the ISM at late times, relative to SNIa enrichment. The C+N enhancement in our two youngest clusters might be a consequence of such extra AGB enrichment, and it is tempting to ascribe the elevated levels of Na, S, and Cu in these clusters to the same source.  However, the theoretically predicted AGB contribution to these three elements is small, in contrast to Ce and C+N (see, e.g., Fig. 10 of \citealt{Andrews2017} and Fig. 13 of \citealt{Rybizki2017}). 

Na overabundance in evolved stars can be caused by mixing during the first dredge-up in low- to intermediate-mass red giants \citep[][and references therein]{Smiljanic2016}. Our observed Na enhancement in only the youngest clusters could reflect this phenomenon since their evolved stars (which make up our sample) are more massive compared to older clusters. In contrast to \citet{Griffith2022}, we do not see any notable Ca or K enhancement in any clusters; however, these enhancements are strongest in clusters younger than $\approx$300 Myr, and we do not have any clusters of this age in our sample. Conversely, \citet{Griffith2022} did not see any enhancement in Na. We also observe mild Cu enhancement in our youngest clusters where they observed depletion. The other elements showing enhancement (S and C+N) were not considered in \citet{Griffith2022}. An advantage of APOGEE for investigating young clusters is that the abundance determinations for red giants are less likely to be affected by rotational line broadening than determinations from main sequence stars that dominate GALAH cluster samples.

\begin{figure*}
    \centering
    \includegraphics[width=0.8\textwidth]{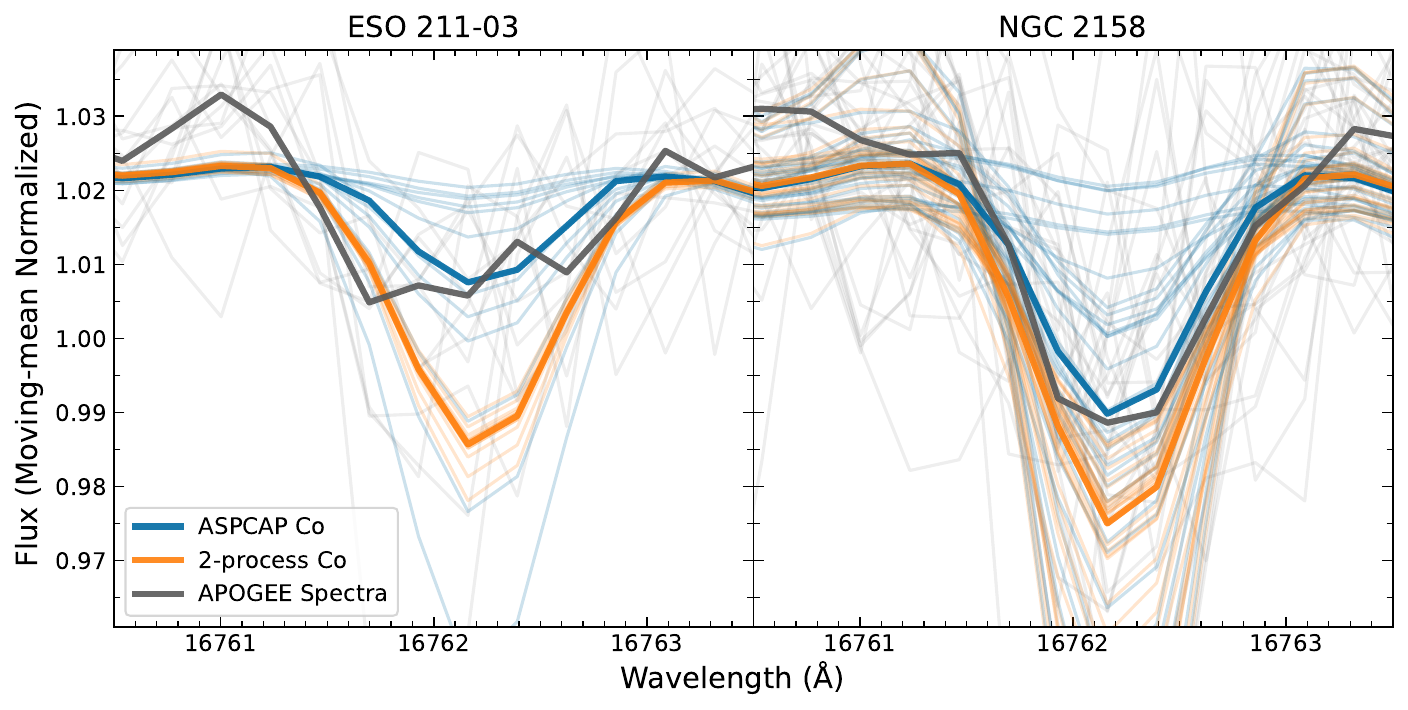}
    \caption{Spectra around the strongest Co absorption line for all member stars in our sample of the open clusters ESO 211-03 and NGC 2158, which showed significant depletion in their median Co residual abundances. This line was identified from the wavelength range where the synthetic spectra, generated with identical abundances \textit{except} Co, showed the largest difference. We note that three other Co lines were identified through this method; we only plot the strongest one here. For each star in the clusters, synthetic spectra were generated with Korg \citep{Wheeler2023a,Wheeler2023b} using the reported ASPCAP abundances only (blue) and two-process predicted Co abundance with ASPCAP abundances for all other elements (orange). Observed APOGEE spectra are in gray. Thin, lighter lines represent individual cluster stars while the thick, dark lines represent the median spectrum. All individual spectra were normalized using a moving mean prior to calculation of the median spectrum; synthetic spectra were downsampled to APOGEE resolution prior to normalization. Better agreement of the observed spectra with the ASPCAP Co abundance indicates that the cluster stars are genuinely depleted in Co relative to the two-process prediction.}
    \label{fig:Co_depletion}
\end{figure*}

ESO 211-03 and NGC 2158 show significant ($>$0.2 dex) depletion in Co. To investigate this depletion, we compare the strongest Co line in the observed APOGEE spectra to synthetic spectra. Using the 1D spectral synthesis code Korg \citep{Wheeler2023a,Wheeler2023b} with the APOGEE DR17 line list (most similar to that of the DR16 line list, described in \citealt{Smith2021}), we generate two synthetic spectra for every star in each cluster: one spectrum with reported ASPCAP abundances and stellar parameters, and one spectrum where only the Co abundance is changed to that predicted by the two-process model. We note that we use the \textit{uncalibrated}, best-fit ASPCAP spectroscopic $\Teff$ and $\log(g)$ as Korg inputs to synthesize the spectra. The synthetic spectra are downsampled to APOGEE resolution, then normalized by dividing out a moving mean, where at each pixel the mean is calculated using a bandwidth of 25\AA. For the observed spectra, we use the combined, but not normalized, APOGEE spectra in the \texttt{apStar}/\texttt{asStar} files and normalize using the same moving mean procedure with a weighted mean to account for flux uncertainties. Finally, we calculate the pixel-by-pixel median of the normalized spectra over every star in the cluster. As illustrated in Figure \ref{fig:Co_depletion}, we find that the median observed spectrum is a closer match to the median synthetic spectrum with reported ASPCAP Co abundances in both ESO 211-03 and NGC 2158. This indicates that the Co prediction from the two-process model results in a poorer fit, and thus that the observed depletion is most likely real. However, the physical origin of this Co depletion is unclear. An alternative explanation for the significant mismatch between the observed spectra and two-process prediction is that the spectra are affected by small errors in sky subtraction: the 16762\AA\ Co line falls on a telluric feature in both ESO 211-03 and NGC 2158. In either case, we show that our median residual abundance analysis can reveal real differences in spectra of stellar populations. A caveat of this analysis is that while we use identical line lists and model atmospheres, Korg is not the same spectral synthesis code used to generate the ASPCAP spectral grids for DR17, so we are in effect generating new spectra to compare to APOGEE.

NGC 1245, NGC 7789, and NGC 2420 show a $\approx$0.1-0.2 dex enhancement in median V in Figure \ref{fig:2proc_OCs_1}. We performed a similar analysis to that of the Co depletion for ASPCAP V (chosen over BAWLAS due to a higher number of measurements) and found that the median observed APOGEE spectra more closely matched that of the median synthetic spectrum with predicted two-process V abundances for all three clusters. From this result we conclude that the V enhancements seen in Figure \ref{fig:2proc_OCs_1} are likely not real because the median line depth of the observed spectra is consistent with the two-process prediction (i.e., the residual should be close to 0). The V residuals show some of the largest variations between individual stars in the majority of the clusters, so there are likely large uncertainties in the measurements as well.

We repeated the median spectrum analysis for the Na enhancement, again using ASPCAP over BAWLAS due to a larger number of measurements, with the two youngest clusters, NGC 6705 and Berkeley 85. However, the continuum level of the median observed spectra did not match that of the synthetic spectra despite identical moving-mean normalization procedures prior to the median calculation, making it difficult to accurately judge the relative depth of the strongest Na lines in each median spectrum. It is therefore unclear whether the median Na enhancement in NGC 6705 and Berkeley 85 is real from this analysis. We note that V and Na are both elements whose lines can become weak at warmer temperatures, and it is possible that some of the V and Na measurements are fitting noise in the spectra in young clusters. This is more likely for V than Na given that the median BAWLAS Na is also elevated in NGC 6705 and Berkeley 85, and significant Na enhancement has been independently reported for NGC 6705 \citep{LoaizaTacuri2023}.

Finally, we check the enhancements for the remaining single elements showing age trends, S, Ce, and Cu, using the median spectrum analysis in Korg. For Ce, we performed this analysis on five clusters showing $>$0.1 dex median enhancement (NGC 6705, NGC 1245, ESO 211-03, NGC 2420, and NGC 2204) and used the ASPCAP abundances to increase the number of measurements in each cluster. The median spectra show clear enhancements in Ce, similarly to the Co case (Figure \ref{fig:Co_depletion}) but to a smaller degree, in all clusters except NGC 6705. NGC 6705 suffered similar continuum matching issues as in the Na case. In S, we found that the median observed spectra were actually shallower than even the two-process prediction i.e., the clusters have less S than predicted. Upon further investigation, it is likely that several line blends in S in combination with the initial fitting of S with the other $\alpha$ elements resulted in ASPCAP overestimating the S abundance. However, in NGC 6705 and Berkeley 85, using the smaller number of BAWLAS measurements of S, whose methods are intended to better measure abundances of elements suffering from blended lines, yields very similar results to using ASPCAP. For Cu, we examined NGC 6705 and Berkeley 85, and we find a similar result to S, where the median observed spectra were shallower than the median synthetic spectra. However, the $\approx$0.1 dex median Cu enhancements seen in these clusters is insufficient to generate a clear difference between the median synthetic spectra generated with measured or predicted (two-process) Cu at APOGEE resolution, so it would have been difficult to determine which is closer to the data regardless.

Overall, we find that open clusters appear increasingly similar to the MW disk as they get older. We find age trends in the elements C+N, Na, S, Cu, and Ce where the two youngest clusters show the substantial enhancements, and all clusters younger than 2 Gyr show enhanced Ce. Detailed examination of spectral fits shows that the Na, S, and Cu enhancements are not fully convincing, as all of them arise from weak features. However, the strong Co depletions in ESO 211-03 and NGC 2158 and the enhancements in Ce in NGC 1245, ESO 211-03, NGC 2420, and NGC 2204 do appear to be real. 

\subsection{Globular Clusters} \label{subsec:GCs}

Once thought to be simple stellar populations, the last several decades have shown that globular clusters (GCs) are complex systems showing evidence of multiple generations of stars. Almost all GCs show star-to-star variation in their light element abundances, an observation that is commonly interpreted as a signature of self-enrichment within the cluster \citep[e.g.,][and references therein]{Gratton2004,Gratton2012}. The most notable chemical signatures include the Na-O anticorrelation and the Al-Mg anticorrelation/Al-Si correlation, which originate from high-temperature proton-capture reactions (i.e., H-burning) only possible in the cores of the first-generation, low-metallicity stars that go on to enrich later generations \citep[e.g.,][]{Carretta2009a,Carretta2010,Carretta2013,Meszaros2020}.

In this section, we examine the stellar populations of GCs through residual abundance analysis. By using residual abundances, we are able to control for the bulk abundance variations resulting from CCSN and SNIa contributions, which are largely shared by field stars in the MW. Thus, this method highlights element variations and correlations potentially unique to GCs, which in turn could be a signature of their distinctive evolutionary histories or interesting nucleosynthetic pathways more common in these systems. Many proposed early-generation polluters of GC stars, such as AGB stars \citep{Ventura2001}, mass loss from fast-rotating massive stars \citep{Decressin2007}, and binary massive stars \citep{deMink2009}, are not necessarily accounted for by our two-process model.

We select GC member stars by cross-matching our residual abundance catalog with the DR17 value-added catalog of Galactic GCs \citep{Schiavon2023}, then impose a stricter set of membership cuts. We require that each star have a membership probability $>$0.8 using either set of independent membership probabilities reported in the catalog. Furthermore, the GC star sample is divided into \textit{Likely} and \textit{Outlier} groups; we use only stars selected using a set of metallicity-independent angular distance, proper motion, and radial velocity criteria in the \textit{Likely} group (\texttt{iFlag}=0) (see \citealt{Schiavon2023} for more details). After these membership cuts, a total of 23 GCs are represented in our sample, and we select eight with $N>8$ stars. M54 was dropped from our GC sample due to heavy contamination from stars in the Sagittarius Dwarf Galaxy (Sgr). We found significant overlap between the \citet{Schiavon2023} M54 sample ($N=313$) and our $N=595$ Sgr sample in Section \ref{subsec:gals}, and it is difficult to separate the M54 and Sgr populations at the metallicity range examined in this work. We choose to examine Sgr as a whole, without separating out any M54 stars, in Section \ref{subsec:gals}. The stars selected in each GC typically span the full giant branch in $\Teff$ and $\log(g)$, except for NGC 6388, NGC 6380, and NGC 6760, whose member stars are all upper RGB ($\log(g)\lesssim1.5$).

\begin{figure*}[!ht]
    \centering
    \includegraphics[width=\linewidth]{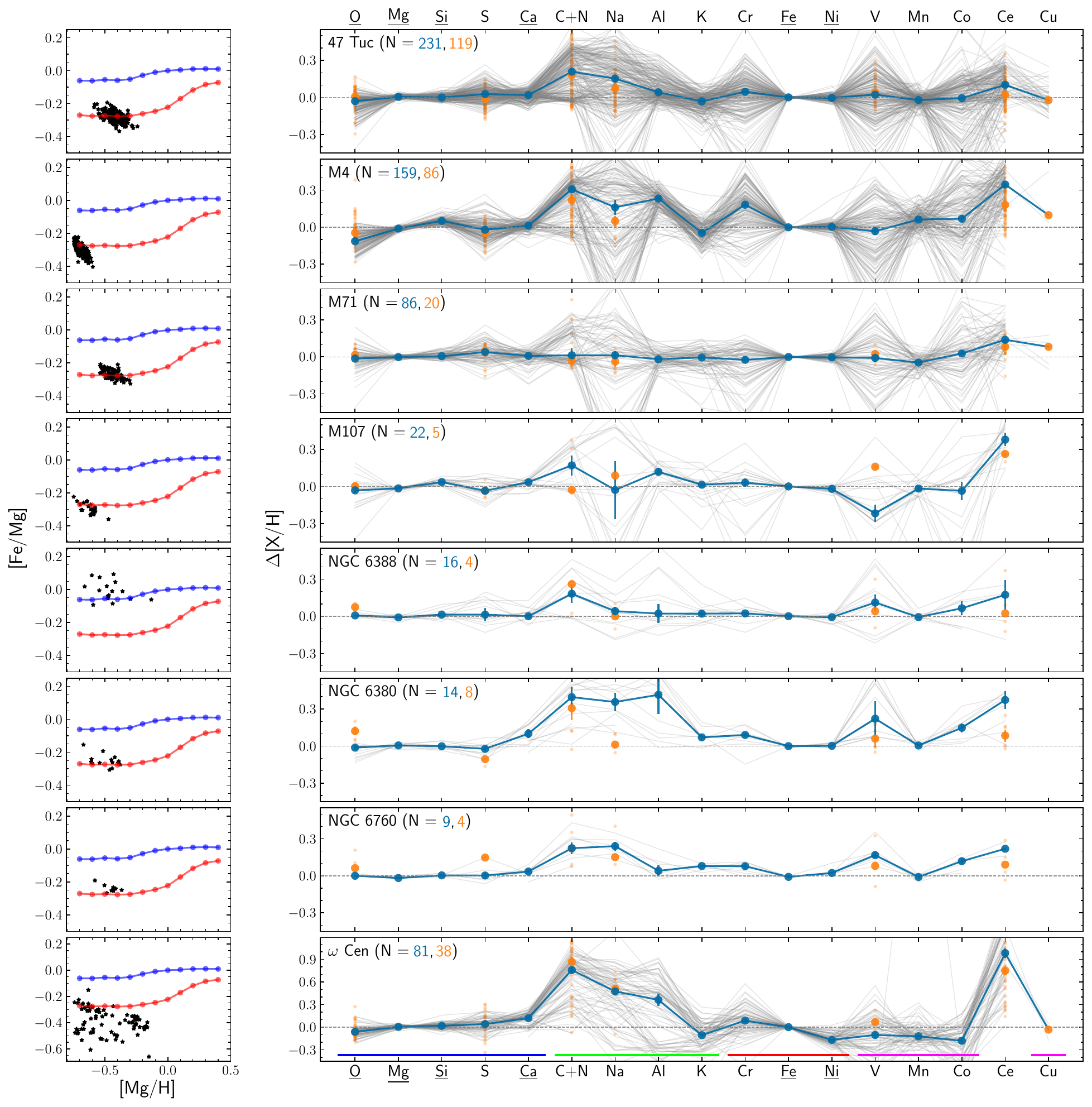}
    \caption{Same as Figure \ref{fig:2proc_OCs_1} for the 8 globular clusters with more than 8 stars in our sample ordered by sample size, except for $\omega$ Cen, which has a different vertical range and is placed last.}
    \label{fig:2proc_GCs}
\end{figure*}

In Figure \ref{fig:2proc_GCs} we plot the residual abundances of our sample of 8 GCs as well as their member stars' location in [Fe/Mg] vs. [Mg/H] space. Because GCs are typically metal-poor but we impose a cutoff of $\text{[Mg/H]}\geq-0.75$ in our sample selection (Section \ref{subsec: sample}), we have ultimately selected relatively metal-rich GCs and/or sampled the metal-rich end of their metallicity distributions. The most metal-poor cluster in our sample is $\omega$ Cen, with $\text{[Fe/H]} = -1.53$ \citep[][2010 edition]{Harris1996}, whose stars have lower SNIa contributions than the median low-Ia plateau, as seen in the left panel of Figure \ref{fig:2proc_GCs}. Our GCs all populate the low [Mg/H] bins, and most also lie on the low-Ia sequence, except for NGC 6388 (a relatively metal-rich bulge cluster).

\begin{figure*}[!ht]
    \centering
    \includegraphics[width=0.9\linewidth]{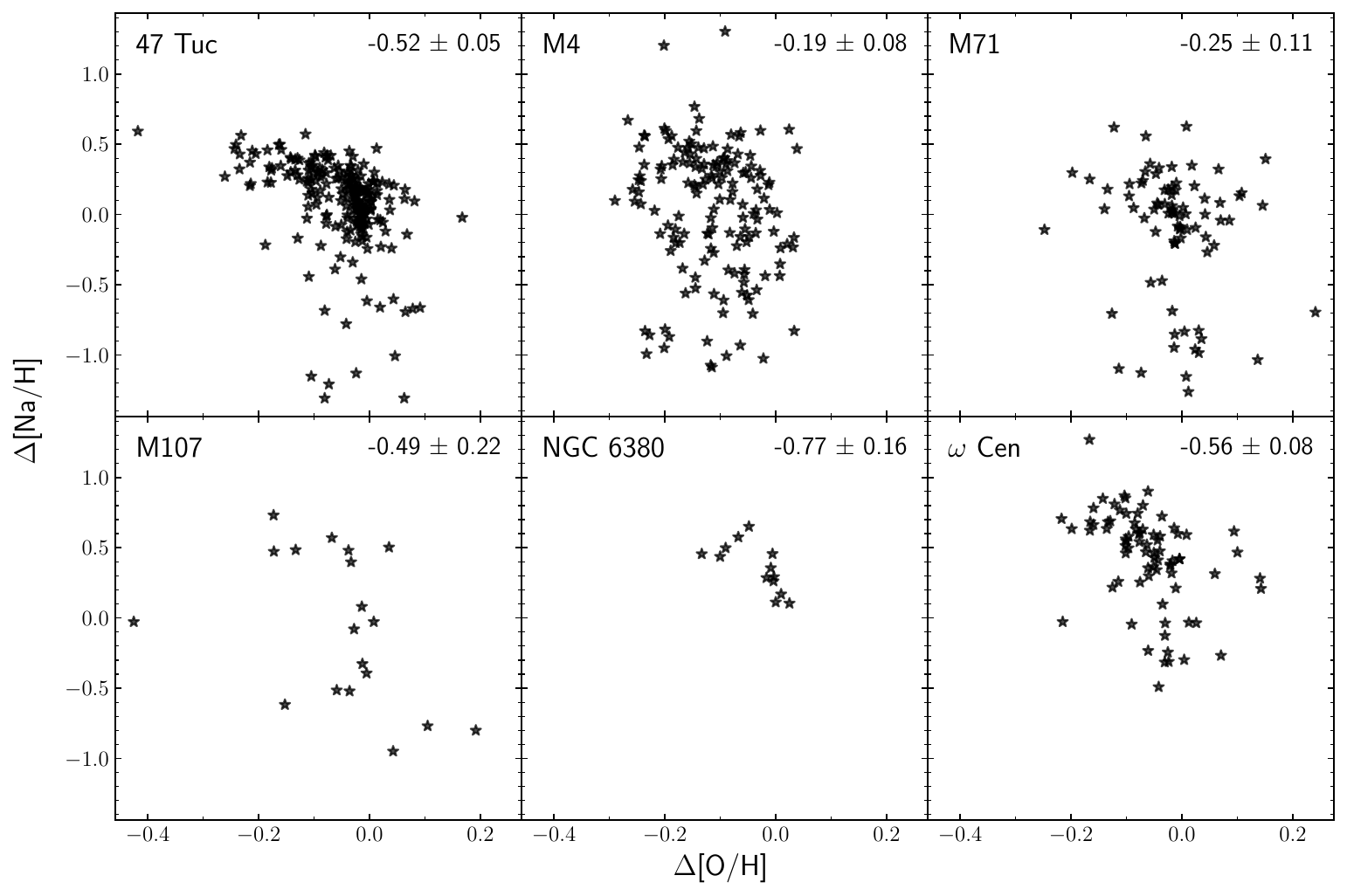}
    \caption{Na-O anticorrelation plot showing residual abundances in ASPCAP [O/H] on the x-axis and ASPCAP [Na/H] on the y-axis for six globular clusters in our sample with statistically significant ($p<0.05$) Spearman correlation coefficients, reported in the upper right of each plot for each cluster. One-$\sigma$ uncertainties are estimated from the standard deviation of the correlation coefficients from 1000 bootstrap resamplings of each cluster.}
    \label{fig:Na-O_anticorrelation}
\end{figure*}

The GCs in our sample all show unique residual abundance patterns. While there are some common trends, none are universal among all clusters in our sample except the very small residuals in Mg, Si, and Fe, which is to be expected since these elements are well-measured in ASPCAP and were used to fit $\Acc$ and $\AIa$. As for the other elements that the two-process model was conditioned on, NGC 6380 and $\omega$ Cen show a $\approx$0.1 dex enhancement in Ca, and $\omega$ Cen shows a $\approx$0.2 dex depletion in Ni. S, K, and Mn also generally show very small median residuals among our GC sample. The only notable K median residuals are $\approx$0.1 dex enhancements in NGC 6380 and NGC 6760 and an $\approx$0.1 dex depletion in $\omega$ Cen; similarly, the only notable Mn residuals are a $\approx$0.1 dex enhancement in M4 and $\approx$0.1 dex depletion in $\omega$ Cen.

\subsubsection{Oxygen and Sodium}\label{subsubsec:GC_O_Na}
The median O value is close to the two-process prediction for all clusters but shows notable star-by-star scatter in most (this effect is most visible in GCs with large $N$). The clusters with the most obvious star-to-star scatter in O also have large star-to-star scatter in Na, which has a known anticorrelation with O in GCs \citep[e.g.,][]{Carretta2009a,Carretta2009b,Carretta2010}. We plot the O and Na residual abundances in Figure \ref{fig:Na-O_anticorrelation} for six GCs in our sample with statistically significant ($p<0.05$) Spearman correlation coefficients. We choose to use the Spearman test because it is more robust to outliers and makes no assumptions about the linearity of the correlation, and we also use residuals from the ASPCAP measurements to maintain a sufficiently large sample size. All six GCs in Figure \ref{fig:Na-O_anticorrelation} have negative Spearman correlation coefficients, indicating that the O and Na residual abundances are indeed anticorrelated. The two clusters with statistically insignificant ($p>0.05$) correlation coefficients, NGC 6388 and NGC 6760, show smaller spreads in $\Delta$[O/H] and/or $\Delta$[Na/H] values compared to the other GCs in Figure \ref{fig:2proc_GCs}. We note that Na has fairly weak lines in the H-band, and previous studies of APOGEE GCs have opted out of studying the Na-O anticorrelation for this reason (e.g., \citealt{Meszaros2020}).

In 47 Tuc, M4, M71, M107, and (to a lesser extent) $\omega$ Cen, we observe a bifurcation in the star-by-star residuals for Na in Figure \ref{fig:2proc_GCs}; this may be a signature of at least two distinct populations within those GCs. The median Na residual would then reflect the relative size of those multiple populations. It is not clear whether these multiple populations exist in NGC 6338, NGC 6380, or NGC 6760 based on their star-to-star spread in Na residuals, although NGC 6380 does show two fairly distinct ``clumps" of stars in $\Delta$[O/H]-$\Delta$[Na/H] space. NGC 6380 and NGC 6760 are also notably the only two GCs with only positive Na residuals.

\begin{figure*}
    \centering
    \includegraphics[width=0.9\linewidth]{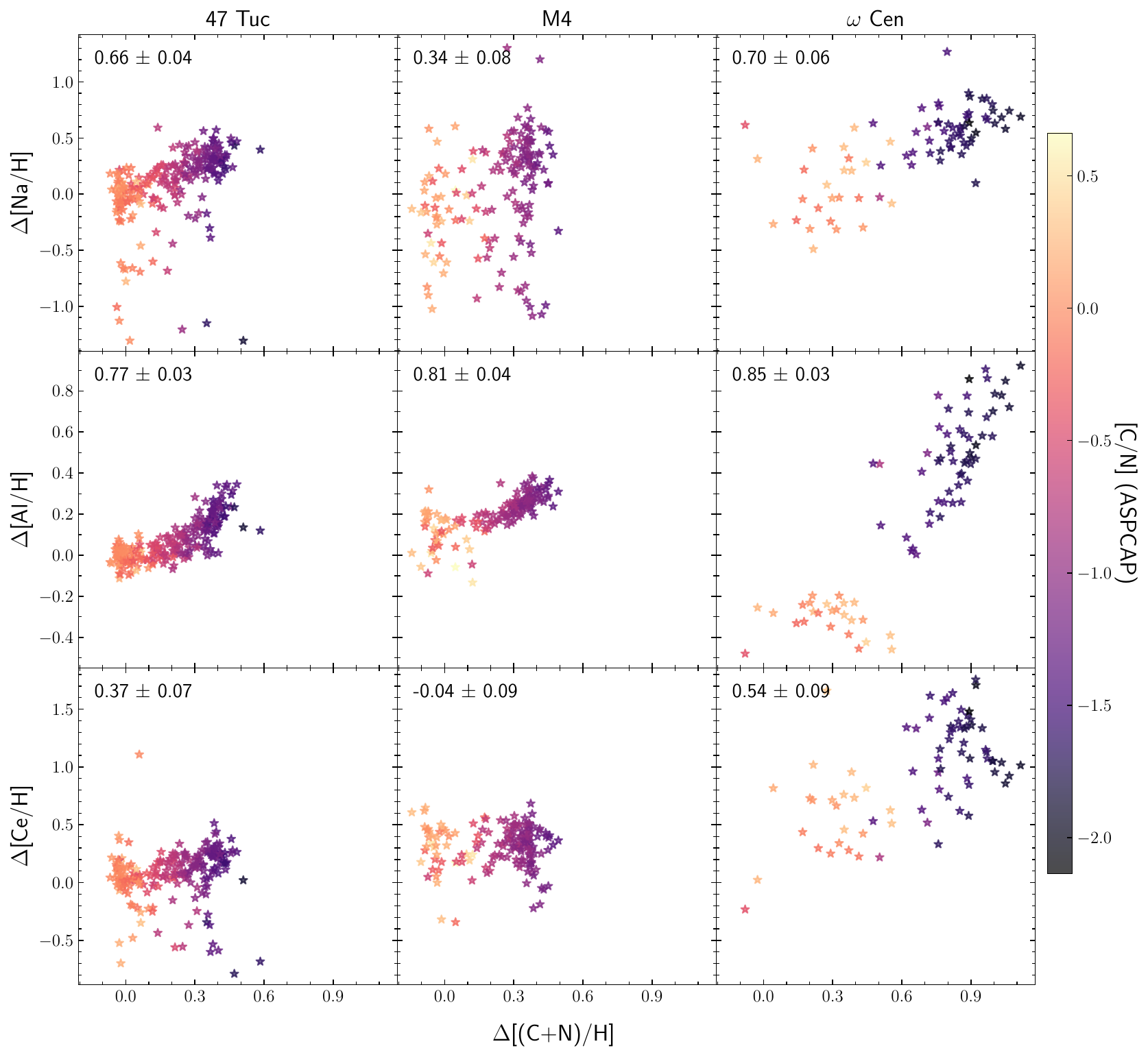}
    \caption{Correlations between residual abundances of ASPCAP [Na/H] (top row), ASPCAP [Al/H] (middle row), and ASPCAP [Ce/H] (bottom row) with ASPCAP [(C+N)/H] for the globular clusters 47 Tuc (left column), M4 (middle column), and $\omega$ Cen (right column). Points are colored by the [C/N] ratio calculated from raw (not adjusted for $C_{\log(g)}$ and $C_{ZP}$) ASPCAP [C/Fe] and [N/Fe] values. Spearman correlation coefficients are reported in the upper left for each cluster, with their 1$\sigma$ uncertainties estimated from the standard deviation of the correlation coefficients from 1000 bootstrap resamplings of each cluster-element combination. All correlation coefficients are statistically significant ($p<0.05$) except in the $\Delta$[Ce/H]-$\Delta$[(C+N)/H] panel for M4.}
    \label{fig:C+N-Na/Al/Ce_correlations}
\end{figure*}

\subsubsection{Carbon, Nitrogen, Aluminum, and Cerium}\label{subsubsec:GCs_C+N_Al_Ce}
All the clusters in our sample show elevated C+N except for M71, which is unusually similar to the MW disk stars compared to the rest of the sample and not significantly enhanced in any element except Ce. The clusters all also show some amount of enhancement in Ce ranging from slight ($\approx$0.1 dex, 47 Tuc and M71) to extreme ($\approx$1 dex, $\omega$ Cen). 47 Tuc and $\omega$ Cen show statistically significant ($p<0.05$) positive correlations between ASPCAP $\Delta$[(C+N)/H] and $\Delta$[Ce/H], plotted in the bottom row of Figure \ref{fig:C+N-Na/Al/Ce_correlations}. NGC 6388 and NGC 6380 also show statistically significant correlations for this pair of elements, but are not plotted. The elevated C+N and Ce abundances could be a signature of second-generation AGB enrichment in these clusters, especially $\omega$ Cen \citep[e.g.,][]{Smith2000}. 

M4, M107, $\omega$ Cen, and NGC 6380 show enhancement in Al. The star-by-star Al residual abundances in $\omega$ Cen (as well as NGC 6388 and NGC 6380, though limited by the much smaller sample sizes) show large, somewhat bimodal scatter around the median, similar to the pattern seen in the Na residuals, and could be a signature of multiple populations within the cluster. However, the star-by-star residual scatter around Al in M4, M71, and M107 is much tighter and appears mostly centered around the median residual value, consistent with the small Al spreads found in \citet{Meszaros2020} for these GCs. An Al-Mg anticorrelation and an Al-Si correlation originating from the Al-Mg cycle in the high temperature ($>$70 million K) cores of early-generation low-metallicity stars has been noted in the literature for many GCs \citep[e.g.,][]{Carretta2009a,Carretta2010,Carretta2013,Meszaros2020}, but we do not see evidence of the Al-Mg relation in our residual abundance data. Five of our GC sample (47 Tuc, M4, NGC 6388, $\omega$ Cen, and M54) have some correlation between Al and Si residuals, but we caution that the spread in $\Delta$[Si/H] is relatively small, only $\approx$0.1-0.2 dex. Mg and Si are two of the six elements used to fit $\Acc$ and $\AIa$ in the two-process model, which leads us to expect little scatter in the Mg and Si residuals.

In the first two rows of Figure \ref{fig:C+N-Na/Al/Ce_correlations}, we plot the residual abundance correlations of Na and Al with C+N for 47 Tuc, M4, and $\omega$ Cen, with points colored by [C/N] ratio. These three clusters are both well represented in our residual abundance catalog ($N>81$) and show notable residuals in the light odd-Z elements and Ce in Figure \ref{fig:2proc_GCs}. While not plotted, all GCs in our sample show similar statistically significant positive correlations between the C+N and Al residuals; NGC 6388 and NGC 6380 show the same between the C+N and Na residuals. 

The strong color gradient along the positive correlations in Figure \ref{fig:C+N-Na/Al/Ce_correlations} shows that [C/N] values decrease as $\Delta$[(C+N)/H] increases, so the C+N enhancement in these GCs is most likely driven by enhancement in N. Considering this, the correlation we observe between Al and C+N residuals is likely a reflection of an Al-N correlation \citep[e.g.,][]{FernandezTrincado2019,Meszaros2020}; similarly, our $\Delta$[Na/H]-$\Delta$[(C+N)/H] correlation likely reflects a Na-N correlation \citep[e.g.,][]{Carretta2009b}. These correlations are, like other known elemental correlations in GCs, believed to be related to pollution from early-generation stars hot enough to burn hydrogen through the Ne-Na and/or Mg-Al cycles. The strong N dependence on the C+N enhancement in these GCs and the correlations with Al and Na are common with many N-enhanced stars found in the MW and have been suggested as evidence for these N-rich stars having GC origin \citep[e.g.,][]{Martell2016,Schiavon2017,FernandezTrinicado2019b}.

\subsubsection{Iron-peak Elements}\label{subsubsec:GCs_FePeak}
The two Fe-peak elements (Fe and Ni) used to fit the two-process amplitudes generally have a very small, if any, median residual in our GC sample except for $\omega$ Cen, which shows a $\approx$0.2 dex depletion in Ni. $\omega$ Cen also has $\approx$0.1-0.2 dex depletions in the other Fe-peak elements V, Mn, and Co in addition to the aforementioned Ni; this chemical pattern in the Fe-peak elements appears more similar to our sample of massive dwarf satellite galaxies (Figure \ref{fig:2proc_dwarfgals}) than the other GCs so we discuss its potential origins in Section \ref{subsubsec:gal_trends}. M4 has an $\approx$0.2 dex median residual in Cr, while NGC 6388 and NGC 6380 show smaller $\approx$0.1 dex Cr residuals. NGC 6388 and NGC 6380 also have $\approx$0.1-0.2 dex enhancements in Co. NGC 6388, NGC 6380, and NGC 6760 show $\approx$0.1-0.2 dex enhancement in V. V also shows large star-by-star scatter in its residuals; however, because the star-by-star deviations are present and large in other stellar populations (e.g. Figures \ref{fig:2proc_OCs_1}, \ref{fig:2proc_OCs_2}, \ref{fig:2proc_dwarfgals}), this is unlikely to be a signature of multiple populations, unlike in the case of Na. Furthermore, M107 shows a V enhancement based on the BAWLAS measurement and a V depletion based on the ASPCAP measurements, which suggests that the variation in V is likely due to large measurement errors. This effect is present but smaller in $\omega$ Cen, where the median BAWLAS residual is just barely positive while the median ASPCAP residual shows a small depletion of $\approx$0.1 dex. Only four GCs in our sample (47 Tuc, M4, M71, and $\omega$ Cen) have any Cu measurements. Of these four GCs, only 47 Tuc has more than six measurements, and it has $\Delta$[Cu/H]$\approx$0.

\subsubsection{Omega Centauri}
The chemical properties of $\omega$ Cen have long been known to be extreme relative to other GCs \citep[e.g.,][]{Johnson2010,Meszaros2021}. From Figure \ref{fig:2proc_GCs}, its residual abundances are much higher than other clusters, but the overall residual abundance pattern---particularly the enhanced light elements C+N, Na, and Al---is actually quite similar to those of the other GCs in our sample (most notably 47 Tuc, M4, and NGC 6380). This greater similarity could be due to sampling only the high metallicity population of our GCs, or that controlling for CCSN and SNIa contributions removes one source of overall abundance variation. Because the residual abundance pattern is similar in form, despite being larger in magnitude, the underlying nucleosynthetic processes creating the abundance patterns (at least for the high-metallicity stars) may be similar.

\subsection{Milky Way Satellite Galaxies} \label{subsec:gals}
Dwarf galaxies in the Local Group are interesting laboratories for probing star formation over a range of masses, metallicities, and environments on a detailed, star-by-star level. The chemical abundance trends of dwarf galaxies provide further insight on the more subtle details affecting star formation, such as variations in the initial mass function \citep[e.g.,][]{McWilliam2013,Hasselquist2017} and AGB enrichment \citep[e.g.,][]{Skuladottir2019,Hansen2018,FernándezTrincado2020}. The low-metallicity environments (compared to the MW disk) of dwarf galaxies could also help identify and constrain metallicity-dependent yields of certain elements.

Our two-process model is calibrated using median abundance trends in the Galactic disk. Therefore, by construction, we can consider the residual abundances in external galaxies as deviations/differences from the chemical pattern of the Galactic disk. Strong residuals (or lack thereof) from the disk-trained two-process model would provide insight into which elements are (or are not) affected by the differences in star formation and chemical enrichment history between the dwarf galaxies and the Galactic disk.

We select our sample of Milky Way satellite galaxies by cross-matching the APOGEE IDs of our sample with Table 2 of \citet{Hasselquist2021}, which examined the chemical abundance trends of the five most massive MW satellites. We find four satellite galaxies represented in our sample: the Large Magellanic Cloud (LMC), Small Magellanic Cloud (SMC), Sagittarius Dwarf Galaxy (Sgr), and Gaia Sausage/Enceladus (GSE). For the LMC and SMC, we also checked for selection effects on our residual abundances by comparing with a sample chosen using a different set of selection criteria, a combination of APOGEE flags and observing fields with proper motion and radial velocity cuts from \citet{Nidever2020}, and found little difference in the residual abundance pattern.

Sgr is currently in the process of merging with the MW. The \citet{Hasselquist2021} Sgr sample contains about two-thirds main body stars. They also demonstrate that despite different metallicity distribution functions between the Sgr core and tidal tails, the [Mg/Fe] vs. [Fe/H] tracks are not significantly different where they overlap in [Fe/H]. Our $\text{[Mg/H]}>-0.75$ metallicity cut (Section \ref{subsec: sample}) selects the highest metallicity stars in Sgr, where the main body stars dominate. Therefore, we do not distinguish between Sgr core and tail stars in this analysis, and we expect the majority of Sgr stars in our sample to be main body stars.

\begin{figure*}[!th]
    \centering
    \includegraphics[width=\linewidth]{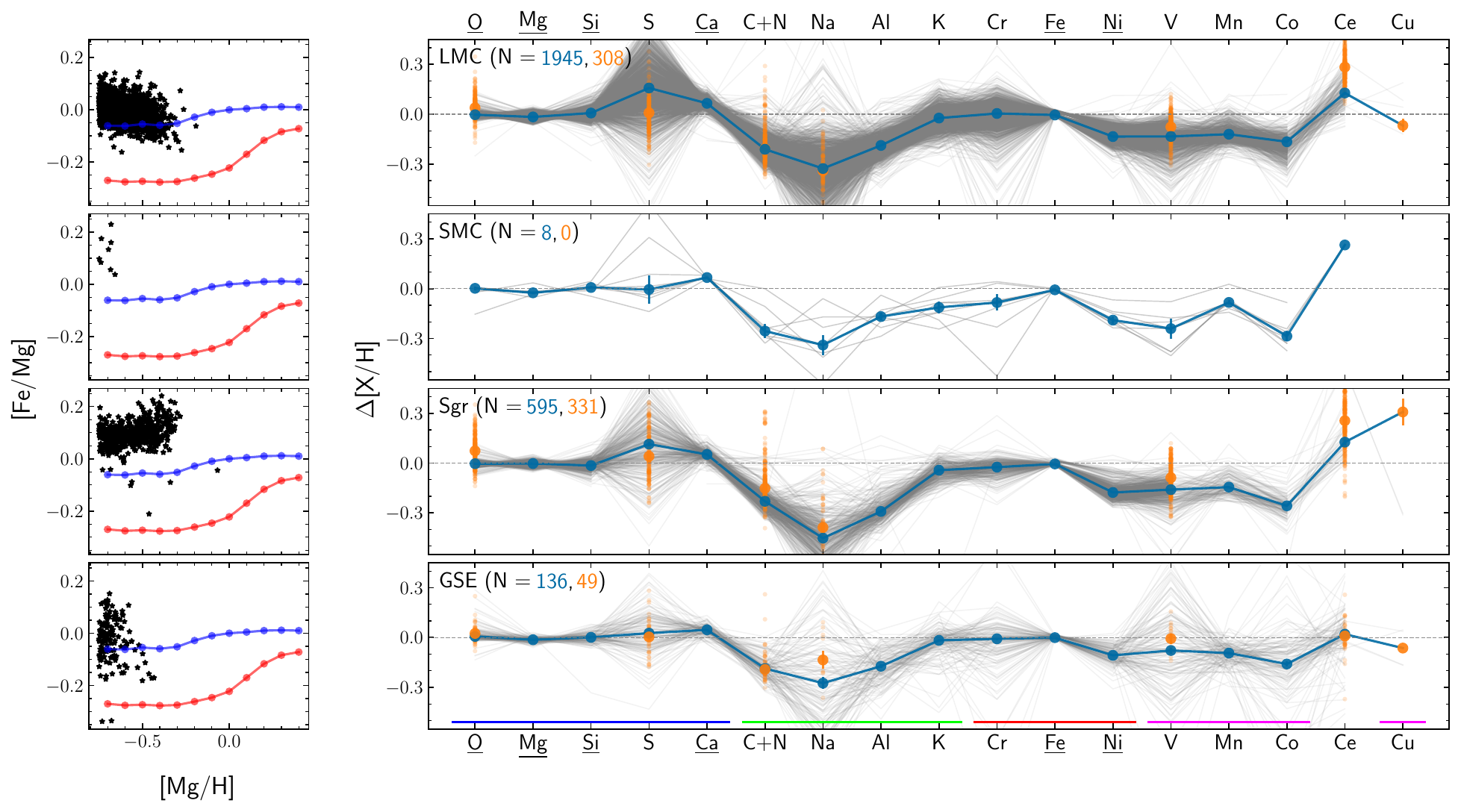}
    \caption{Same as Figure \ref{fig:2proc_OCs_1} for four massive MW satellite galaxies: the LMC, SMC, Sgr, and GSE.}
    \label{fig:2proc_dwarfgals}
\end{figure*}

We plot [Fe/Mg] vs. [Mg/H] for the member stars in each of the satellite galaxies in our sample in the left panels of Figure \ref{fig:2proc_dwarfgals}. All four galaxies lie on or above the median high-Ia sequence, indicating extended star formation histories: the gas that formed the most metal-rich stars in these systems had time to be enriched by many generations of delayed SNIa, thus increasing the relative amount of Fe to Mg. GSE shows the greatest number of stars below the high-Ia median sequence, which makes sense given that its star formation was most likely cut off $\sim$10 Gyr ago during its infall into the MW \citep[e.g.,][]{Helmi2018,Gallart2019,Vincenzo2019}, so later SNIa contributions were not incorporated into the most metal-rich stars. A ``wall" of stars exists at the leftmost edge of the panels, right at the $\text{[Mg/H]}\geq-0.75$ sample metallicity cutoff. The metallicity range we can probe in this study is limited by the metallicity range of the MW disk calibration sample, chosen to match \citetalias{Weinberg2022} and such that we observe clear [$\alpha$/Fe] bimodality over its metallicity range. When compared to the [Fe/Mg] vs. [Mg/H] plots in \citet{Hasselquist2021} (Figures 6 and 7), it is clear that our work samples only the most metal-rich end of these satellite galaxies' metallicity distributions.

On the right panels of Figure \ref{fig:2proc_dwarfgals}, we plot the residual abundances of the stars in our dwarf galaxy sample. Overall, the most striking result are the significantly lower abundances ($\approx$0.2-0.3 dex below the two-process predictions, i.e., the median MW disk trends at matched Fe and Mg), in all light, odd-Z elements except K (which is consistent with the two-process prediction within $\leq$0.1 dex), shared by all four satellite galaxies. The low C+N and Al abundances are consistent with their [X/Mg] tracks in \citet{Hasselquist2021}, which both fall significantly below the MW trends where they overlap in [Mg/H]. We see a similar, though less extreme, $\approx$0.1-0.2 dex depletion in all four satellites for the odd-Z Fe-peak elements, excepting Cu. In both of these element groups, the one exception to the trend is the most massive element, K in the case of the light odd-Z elements and Cu in the case of the odd-Z Fe-peak elements. The $\alpha$-elements O, Mg, and Si show close agreement with median disk trends along with Fe ($\Delta$[X/H]$\approx$0). While this could be a result of using these elements to fit for $\AIa$ and $\Acc$ in the two-process model, we have checked that the median residuals change by $\leq 0.02$ dex if $\AIa$ and $\Acc$ are estimated from Mg and Fe alone.

\subsubsection{Large and Small Magellanic Clouds}\label{subsubsec:MCs}
Our median LMC residual abundances are broadly in agreement with those of the much smaller sample in \citetalias{Weinberg2022}. There is some enhancement in the ASPCAP measurements of the more massive $\alpha$ elements S ($\approx$0.2 dex) and Ca ($\approx$0.1 dex), which is slightly higher than was seen in \citetalias{Weinberg2022}. However, ASPCAP S shows signficant star-to-star scatter and the median BAWLAS S residual is close to 0. The Ca enhancement is consistent with the [Ca/Mg] track for the LMC in \citet{Hasselquist2021}, which is slightly elevated above the MW trend at higher [Mg/H]. K has a median residual of $\approx$0 dex, indicating a high degree of similarity with the MW disk. This strongly contrasts its lighter odd-Z element counterparts C+N, Na, and Al, which show significant depletion compared to the disk. The contrast between the more mass-dependent Na and Al yields and K has been suggested by \citet{Hasselquist2017} to be a signature of a top-light IMF in Sgr; the same argument may hold for the LMC given the similar abundance pattern. However, \citet{Hasselquist2021} show that a top-light IMF may not be required to fit observed abundance patterns, and there is little evidence of such a top-light IMF in the $\alpha$ abundances at metallicities lower than those included in our sample \citep[e.g.,][]{Hansen2018}. Ni shows $\approx$0.1 dex depletion in the LMC, in contrast with lighter even-Z Fe-peak elements Cr and Fe, which are both well-predicted by the two-process model. We also see enhancement in Ce of $\approx$0.1 dex for the ASPCAP measurements and $\approx$0.3 dex for the BAWLAS measurement; for comparison, \citetalias{Weinberg2022} observed a $\approx$0.2 dex Ce enhancement in the LMC. The Ce enhancement is consistent with known enhancements in its fellow second-peak \textit{s}-process elements Ba and La, which may be a signature of increased (compared to the MW) metal-poor AGB enrichment in the LMC's history \citep{VanderSwaelmen2013}. The Ni depletion and Ce enhancement are also consistent with the tracks in \citet{Hasselquist2021}. The median Cu in the LMC is slightly depleted, similar to the other odd-Z Fe-peak elements.

Despite only having 8 stars overlap with our sample, the SMC still shows a distinct residual abundance pattern. It shows strong depletion in the same light odd-Z elements as the other galaxies, and notably is the only galaxy in our sample to show depletion in K ($\approx$0.1 dex). Its depletions in V and Co are also somewhat stronger ($\approx$0.2-0.3 dex) than in the other galaxies. The SMC also shows a very strong enhancement in Ce ($\approx$0.3 dex). We caution that only two stars out of the total eight had an ASPCAP Ce measurement, but this strong enhancement would be roughly consistent if we extrapolated the [Ce/Mg] trend for the SMC in \citet{Hasselquist2021} to higher [Mg/H], where it would fall above the distribution in the MW disk.

\subsubsection{Sagittarius Dwarf Galaxy}\label{subsubsec:Sgr}
The Sgr dwarf galaxy shows a similar residual abundance pattern to the other galaxies in our sample. It has large star-to-star scatter in S (similar to the LMC), but its median ASPCAP S is $\approx$0.1 dex lower than that of the LMC. The Ce enhancement is also similar to the LMC in both its magnitude and the difference between median ASPCAP and BAWLAS measurements. The Sgr Ni track is $\approx$0.1 dex higher than the LMC at high [Mg/H] in \citet{Hasselquist2021}, but the two galaxies show very similar median $\Delta$[Ni/H] in Figure \ref{fig:2proc_dwarfgals}. This suggests that the difference in Ni tracks between Sgr and the LMC in \citet{Hasselquist2021} can be largely explained by differences in CCSN vs. SNIa contribution, as illustrated by the difference in their \citet{Hasselquist2021} [Fe/Mg] tracks and position of member stars in the left panels of Figure \ref{fig:2proc_dwarfgals} relative to the high-Ia median sequence. Sgr shows the strongest Na and Al depletion ($\approx$0.4 and $\approx$0.3 dex, respectively) in our sample, consistent with previous work noting strong deficiencies in these elements in Sgr stars \citep[e.g.,][]{Bonifacio2000,Sbordone2007,McWilliam2013,Hasselquist2017}. Additionally, the lack of K depletion is consistent with a top-light IMF scenario in Sgr proposed by \citet{McWilliam2013} and \citet{Hasselquist2017}, though we again note that this scenario has dubious evidence at metallicities lower than our sample \citep{Hansen2018}. The $\approx$0.2 dex C+N depletion is about the same as the other galaxies while the $\approx$0.3 dex Co depletion is more similar to that of the SMC than the LMC or GSE. The most notable feature of Sgr's residual abundance pattern is its high ($\approx$0.3 dex) median Cu enhancement, derived from residual abundances of 44 stars, which is in disagreement with previously observed Cu deficiencies \citep[e.g.,][]{Sbordone2007,McWilliam2013}. This could be an artifact of a more general disagreement where the Cu abundances in the entire BAWLAS sample are significantly higher than literature abundances at low metallicities ($\text{[Fe/H]}\lesssim-0.8$). The difference is similar in magnitude to our observed $\approx$0.3 dex median Cu residual abundance. \citet{Hayes2022} attribute the discrepancy to a combination of weak, blended lines at APOGEE wavelengths leading to Cu measurements that are biased high at lower metallicity, and NLTE effects that are stronger in optical wavelengths than infrared and should push the optical abundances to larger values.

\subsubsection{Gaia Sausage/Enceladus}\label{subsubsec:GSE}
Like the other dwarf galaxies in our sample, GSE is also depleted in the light odd-Z elements C+N, Na, and Al. It has the smallest Na depletion and largest ASPCAP/BAWLAS discrepancy in our sample, $\approx$0.3 dex using ASPCAP measurements and only $\approx$0.15 dex using BAWLAS measurements. GSE also shows no notable deviations in median Ce, unlike the other galaxies in our sample which show Ce enhancements of $\geq$0.1 dex compared to the disk stars; this is generally consistent with the [Ce/Mg] tracks in \citet{Hasselquist2021}, which show GSE having $\approx$0.1 dex lower [Ce/Mg] than the other dwarfs at the same [Mg/H]. \citet{Hasselquist2021} suggest that the MCs and Sgr had more AGB contributions that led to their Ce enhancements compared to the MW disk, but GSE's shorter star formation time (due to its infall into the MW $\sim$10 Gyr ago) reduced the AGB contribution in its stars. Cu is slightly depleted in GSE, though to a lesser extent than the other odd-Z Fe-peak elements; this pattern is closest to that of the LMC. 

\subsubsection{Common Trends}\label{subsubsec:gal_trends}
\citet{Hasselquist2021} note and discuss the abundances of Al and Ni compared to the MW in all satellite galaxies in our sample; they invoke metallicity-dependent yields and a greater number of metal-poor SNe in these systems than the MW to explain the observed depletions. Since C+N and Na show a similar residual pattern to Al and are all theorized to be ejected from CCSN, it is possible that these elements share similar metallicity-dependent CCSN yields. The lack of depletion in K compared to Na and Al has been cited as evidence for a top-light IMF that produces fewer massive CCSN progenitors in Sgr \citep{McWilliam2013,Hasselquist2017}. Since this chemical signature is present in all four dwarf galaxies we examine (especially the LMC, Sgr, and GSE), this top-light IMF scenario is not unique. However, the top-light IMF is not necessarily required to reproduce overall abundance patterns \citep{Hasselquist2021} and lacks supporting evidence in very metal-poor Sgr stars \citet{Hansen2018}. We also observe that the odd-Z Fe-peak elements V, Mn, and Co share similar depletion magnitudes to Ni; all of these Fe-peak elements are known to have metallicity-dependent yields \citep[e.g.,][]{Woosley1995,Chieffi2004,Nomoto2006}. Furthermore, Ni and Mn depletions have been suggested as potential markers of enrichment from sub-Chandrasekhar SNIa in dwarf galaxies \citep[e.g.,][]{McWilliam2018,Kirby2019,delosReyes2020}, but these previous studies have focused on dwarfs of lower mass and metallicity than we examine here.

In general, low C+N, Na, and Al relative to MW field stars are the strongest chemical signatures of dwarf galaxy populations, and Ni, Mn, and perhaps Co are weaker but also potentially useful tracers. On the other hand, near-zero median $\alpha$-element residuals indicate that dwarf galaxy stars are chemically similar to MW field stars in these elements at fixed metallicity, and thus $\alpha$ abundances \textit{alone} may not be a robust identifier of dwarf populations.

The residual abundance diagram (Figure \ref{fig:2proc_dwarfgals}) provides a compact way to compare and contrast multi-dimensional enrichment patterns among satellites and between satellites and the disk, controlling for differences associated with the overall metallicity and SNIa/CCSN ratio, and leveraging the ability of median statistics to reveal systematic differences below the level of individual star measurement errors. Overall, we find that the four massive MW satellite galaxies represented in our data (LMC, SMC, Sgr, and GSE) show surprisingly similar residual abundance patterns despite their different star formation environments and enrichment histories, as well as only sampling from the most metal-rich stars in these systems due to our sample selection constraints. Combining the $K$-process model (KPM) methodology of \citet{Griffith2023} with the $\log(g)$ calibration strategy developed here would allow these comparisons to be extended to the lower metallicity populations of dwarf satellites, tidal streams, and the in-situ halo.

In forthcoming work, Hasselquist et al. (in prep.) examine two-process fits and residual abundances in dwarf satellites using APOGEE DR17. Although there are many differences of methodology---including the use of matched-$\log(g)$ samples rather than the $\log(g)$ abundance corrections employed here---their results for stars with $\text{[Mg/H]}>-0.75$ are in excellent agreement with those shown in Figure \ref{fig:2proc_dwarfgals}, demonstrating the robustness of these conclusions to analysis details.

\section{Summary} \label{sec:conclusions}

We present a method of correcting systematic abundance trends with surface gravity $\log(g)$ by deriving corrective offsets from median [X/Mg] vs. [Mg/H] abundance trends, which are universal throughout the MW disk and bulge and thus assumed to be independent of stellar evolutionary state. We apply these offsets to 310,427 spectra (288,789 unique stars) from APOGEE DR17. We derive two-process residual abundances for these stars following the method described in \citetalias{Weinberg2022}. Our sample shares a metallicity range with \citetalias{Weinberg2022}, $-0.75\geq$ [Mg/H] $\geq0.45$, but it spans the entire red giant branch and red clump sample in APOGEE: $0\geq\log(g)\geq3.5$ and 3000 K $\geq\Teff\geq$ 5500 K. We consider the $\alpha$-elements O, Mg, Si, S, and Ca; the light odd-Z elements Na, Al, K, and the combined element C+N; Fe-peak elements V, Cr, Mn, Fe, Co, Ni, and Cu; and the \textit{s}-process element Ce. Abundances of all elements except Cu are from the APOGEE abundance measurement pipeline ASPCAP \citep{Holtzman2015,GarcíaPérez2016}. The Cu abundances, analyzed in context of a two-process model and residual abundances for the first time in this work, are taken from the BAWLAS catalog \citep{Hayes2022}, along with a set of independent measurements of C+N, O, Na, S, V, and Ce. Our calibration offsets and two-process residual catalog are available online (see Appendices).

As a first application of our expanded residual abundance catalog, we examine the median residual abundance trends of 14 open clusters, nine globular clusters, and four MW dwarf satellites. Many APOGEE targets in these populations are only accessible through our wide $\log(g)$ and $\Teff$ cuts. Using median abundances reduces the impact of an individual star's measurement errors and reveals systematic differences below the level of those measurement errors. We summarize our key findings below:
\begin{enumerate}
    \item Overall, older open clusters show smaller median residual abundances, i.e., they are more chemically similar to MW disk stars. The two youngest clusters show noticeable enhancement in median C+N, Na, S, and Cu compared to older clusters. Ce enhancement is generally seen in clusters with ages up to $\approx$2 Gyr, similarly to young field stars. The Ce and C+N enhancements are likely a signature of enhanced AGB enrichment.
    \item We investigate enhancements or depletions in open clusters through spectral synthesis with Korg \citep{Wheeler2023a,Wheeler2023b}, which has not previously been performed with APOGEE spectra. We find spectral evidence of Co depletion in ESO 211-03 ($\approx$0.6 dex) and NGC 2158 ($\approx$0.25 dex) through this method, although the physical origin of this depletion is unclear.
    \item We recover the well-known Na-O anticorrelation in 6 out of 9 globular clusters in our sample in $\Delta$[Na/H] and $\Delta$[O/H] residual abundance space. Our result indicates that the Na-O anticorrelation is not explained by changes in bulk CCSN/SNIa contributions, and it also highlights the power of residual abundance correlations in identifying unique enrichment pathways.
    \item We find $>$0.2 dex C+N enhancements, and statistically significant correlations in the residual abundances between C+N and Ce, Al, and Na, for several globular clusters in our sample. The (C+N)-Ce correlation is likely a signature of AGB enrichment, while the (C+N)-Al and (C+N)-Na correlations are likely reflections of known Al-N and Na-N correlations. Because our two-process residuals essentially compare populations to median MW disk trends, this result supports claims that N-enhanced stars are of globular cluster origin.
    \item The globular cluster $\omega$ Cen, which is known to have extremely complex and unusual abundance patterns, shows enhanced median \textit{residual} abundances in C+N, Na, and Al similar to that of 47 Tuc, M4, and NGC 6380. While the magnitude of the residual abundances in $\omega$ Cen is larger, the similar chemical abundance pattern suggests similar enrichment mechanisms in the metal-rich populations of these four clusters.
    \item The individual abundance patterns of four of the most massive MW satellites are broadly consistent with previous results comparing these systems to the MW disk population \citep[e.g.,][]{Hasselquist2021}. Our analysis finds surprisingly similar residual abundance patterns among all four satellites once bulk changes in metallicity and CCSN/SNIa contributions are accounted for by the two-process model, despite their different star formation environments and enrichment histories.
    \item Strong Al and Na depletions are observed across the median residual abundances of \textit{all} massive satellite galaxies in our sample. We also observe a milder depletion in Ni of similar strength across our sample. Low Al and Ni abundances are potential signatures of metallicity-dependent SN yields/more enrichment from metal-poor SNe \citep{Hasselquist2021}.
\end{enumerate}

Our expanded two-process residual abundance catalog, which contains $\approx$8 times more stars than previous work, is only possible with the derivation of $\log(g)$ calibration offsets that remove artificial abundance trends and correlations between elements. While we provide values of the $\log(g)$ calibration offsets for our specific sample constraints, our method of empirically calibrating $\log(g)$ trends is generally applicable to any data set, provided one has sufficient stars in each ($\log(g)$, [Mg/H]) bin to derive a robust median value. An immediate usage example that follows from the analysis presented in this work is to use our $\log(g)$ calibration method in the derivation of residual abundances at low metallicities using the KPM method \citep{Griffith2023}, which will in turn enable further comparative studies of populations in dwarf satellites, tidal streams, and the MW halo at lower metallicities than our sample here.

There are many further applications of our expanded residual abundance catalog beyond the comparative study of stellar populations presented in this work. A systematic exploration of outlier stars with large two-process $\chi^2$ fit values should reveal both stars with unusual abundance patterns, which could be signatures of rare physical processes or other enrichment sources, and stars with rare measurement errors. The method of comparing observed spectra to synthetic spectra with the two-process prediction, used in our open cluster analysis here, can help distinguish which deviations are real. Unresolved binaries may be poorly fit by the two-process model and stand out as high $\chi^2$ stars, and trends in residual abundances with, for example, radial velocity variations, could point to chemical signatures of binarity. Furthermore, our expanded sample allows for much greater overlap with external catalogs such as APOKASC \citep{Pinsonneault2018} and \textit{Gaia} DR3 \citep{Gaia2023}, enabling analyses of residual abundance trends with stellar age, kinematics, position in the Galaxy, and more. Furthermore, we can quantitatively test aspects of galactic chemical evolution models: the intrinsic scatter in residual abundances provides constraints on stochastic effects, and the $q_{\text{cc}}$ and $q_{\text{Ia}}$ process vectors provide information about population-averaged yields and their metallicity dependence.

As future spectroscopic surveys, notably \textit{Milky Way Mapper} \citep{Kollmeier2017}, continue to map out increasingly more multi-element abundances across the MW and its dwarf satellite neighbors, it becomes imperative to effectively interpret and analyze the influx of data. As discussed in this paper and previous works (e.g., \citetalias{Weinberg2022}; \citealt{Griffith2022,Griffith2023}) a two-process or K-process model and residual abundances provide a way to reduce/recast high-dimensional abundance information and disentangle the enrichment histories of the Galaxy, its satellites, or special stellar populations from the effects of nucleosynthetic yields. While survey abundance measurement pipelines will continue to improve, the $\log(g)$ calibration method outlined in this work can account for remaining systematics and thus maximize the number of stars on which we can apply these analyses. In turn, we will be rewarded by increasing our understanding of the astrophysical origins of the elements and the processes governing the chemical enrichment of our Galaxy and nearest neighbors.

\section{Acknowledgements}
We thank Emily Griffith and Yuan-Sen Ting for helpful comments on an earlier draft of this paper.

This work is supported by NSF grants AST-1909841 and AST-2307621. TM acknowledges financial support from the Spanish Ministry of Science and Innovation (MICINN) through the Spanish State Research Agency, under the Severo Ochoa Program 2020-2023 (CEX2019-000920-S).

Funding for the Sloan Digital Sky 
Survey IV has been provided by the 
Alfred P. Sloan Foundation, the U.S. 
Department of Energy Office of 
Science, and the Participating 
Institutions. 

SDSS-IV acknowledges support and 
resources from the Center for High 
Performance Computing  at the 
University of Utah. The SDSS 
website is www.sdss4.org.

SDSS-IV is managed by the 
Astrophysical Research Consortium 
for the Participating Institutions 
of the SDSS Collaboration including 
the Brazilian Participation Group, 
the Carnegie Institution for Science, 
Carnegie Mellon University, Center for 
Astrophysics | Harvard \& 
Smithsonian, the Chilean Participation 
Group, the French Participation Group, 
Instituto de Astrof\'isica de 
Canarias, The Johns Hopkins 
University, Kavli Institute for the 
Physics and Mathematics of the 
Universe (IPMU) / University of 
Tokyo, the Korean Participation Group, 
Lawrence Berkeley National Laboratory, 
Leibniz Institut f\"ur Astrophysik 
Potsdam (AIP),  Max-Planck-Institut 
f\"ur Astronomie (MPIA Heidelberg), 
Max-Planck-Institut f\"ur 
Astrophysik (MPA Garching), 
Max-Planck-Institut f\"ur 
Extraterrestrische Physik (MPE), 
National Astronomical Observatories of 
China, New Mexico State University, 
New York University, University of 
Notre Dame, Observat\'ario 
Nacional / MCTI, The Ohio State 
University, Pennsylvania State 
University, Shanghai 
Astronomical Observatory, United 
Kingdom Participation Group, 
Universidad Nacional Aut\'onoma 
de M\'exico, University of Arizona, 
University of Colorado Boulder, 
University of Oxford, University of 
Portsmouth, University of Utah, 
University of Virginia, University 
of Washington, University of 
Wisconsin, Vanderbilt University, 
and Yale University.

\software{
Matplotlib \citep{matplotlib}, astropy \citep{astropy2013,astropy2018,astropy2022}, NumPy \citep{numpy}, SciPy \citep{scipy}, Korg \citep{Wheeler2023a,Wheeler2023b}
}

\appendix
\section{Calibration Offsets} \label{appx:calib_offsets}
All calibration offsets, including the grid of $C_{\log(g)}$ values and values of $C_{\text{ZP}}$ are provided in a table, available online at \url{https://doi.org/10.5281/zenodo.10659205}. The first five rows are shown in Table \ref{table:offsets}. We also include a Jupyter notebook with sample code to load the table, extract the offset grid for a specific element, and obtain calibrated abundances for ten example stars. Also included in the notebook is code to reproduce Figure \ref{fig:logg_offsets}. Note that the reported calibration offsets are derived for and thus applicable to only the data used in this work (described in detail in Section \ref{subsec: sample}).

\begin{splitdeluxetable*}{ccccBcccccccc}
\tablecaption{Table of Calibration Offsets}
\label{table:offsets}
\tablehead{\colhead{Element} & \colhead{Source} & \colhead{C\_ZP} & \colhead{MG\_H} & \colhead{C\_LOGG\_0.25} & \colhead{C\_LOGG\_0.75} & \colhead{C\_LOGG\_1.25} & \colhead{C\_LOGG\_1.75} & \colhead{C\_LOGG\_2.25} & \colhead{C\_LOGG\_2.75} & \colhead{C\_LOGG\_3.25} & \colhead{C\_LOGG\_RC}}
\startdata
O & ASPCAP & -0.025 & -0.7 & 0.122 & 0.110 & 0.077 & 0.009 & -0.079 & -0.081 & -0.081 & -0.058 \\
O & ASPCAP & -0.025 & -0.6 & 0.077 & 0.072 & 0.062 & 0.009 & -0.069 & -0.090 & -0.090 & -0.058 \\
O & ASPCAP & -0.025 & -0.5 & 0.060 & 0.044 & 0.042 & 0.009 & -0.060 & -0.074 & -0.062 & -0.058 \\
O & ASPCAP & -0.025 & -0.4 & 0.043 & 0.024 & 0.030 & 0.009 & -0.037 & -0.050 & -0.056 & -0.034 \\
O & ASPCAP & -0.025 & -0.3 & 0.001 & 0.004 & 0.019 & 0.009 & -0.020 & -0.023 & -0.032 & -0.019 \\
\enddata
\tablecomments{Full table available online at \url{https://doi.org/10.5281/zenodo.10659205}.}
\end{splitdeluxetable*}

\section{Process Vector Tables}\label{appx:q_vals}
In Table \ref{table:qcc} we report values of the CCSN process vector $q^X_{\text{cc}}(z)$ for all elements considered in this work. The same for the SNIa process vector $q^X_{\text{Ia}}(z)$ is reported in Table \ref{table:qIa}. Values are calculated from the median sequences of $\log(g)$-calibrated abundances following Equations 25 and 26 in \citetalias{Weinberg2022}. For more details, see Section \ref{subsec:2proc_vectors}.

\section{Catalog User Guide}
\label{appx:catalog}
The residual abundance catalog is available online at \url{https://doi.org/10.5281/zenodo.10659205}. The table columns are reported in Table \ref{table:catalog_cols}. We also include a Jupyter notebook with sample code to cross-match with external catalogs, using AstroNN \citep{LeungBovy2019a,LeungBovy2019b} and the DR17 globular cluster catalog \citep{Schiavon2023} as examples. The notebook also includes code to make some plots, including reproducing the residual abundances (right side) of Figure \ref{fig:2proc_GCs} and plotting $\AIa/\Acc$ as function of position.

\begin{deluxetable*}{cccccccccccccc}
\tablecaption{Values of $q_{\text{cc}}^X$} \label{table:qcc}
\tablehead{\colhead{Element} & \colhead{Source} & \multicolumn{12}{c}{[Mg/H]}\\
\colhead{} & \colhead{} & \colhead{-0.7} & \colhead{-0.6} & \colhead{-0.5} & \colhead{-0.4} & \colhead{-0.3} & \colhead{-0.2} & \colhead{-0.1} & \colhead{0.0} & \colhead{0.1} & \colhead{0.2} & \colhead{0.3} & \colhead{0.4}}
\startdata
O & ASPCAP & 1.129 & 1.047 & 1.014 & 0.992 & 0.974 & 0.957 & 0.952 & 0.953 & 0.943 & 0.916 & 0.877 & 0.862 \\
O & BAWLAS & 1.522 & 1.491 & 1.363 & 1.278 & 1.201 & 1.142 & 1.076 & 1.005 & 0.919 & 0.800 & 0.760 & 0.660 \\
Mg & ASPCAP & 1.000 & 1.000 & 1.000 & 1.000 & 1.000 & 1.000 & 1.000 & 1.000 & 1.000 & 1.000 & 1.000 & 1.000 \\
Si & ASPCAP & 1.036 & 0.976 & 0.925 & 0.892 & 0.868 & 0.844 & 0.831 & 0.819 & 0.792 & 0.747 & 0.697 & 0.682 \\
S & ASPCAP & 1.153 & 1.152 & 1.099 & 1.066 & 1.047 & 1.003 & 0.961 & 0.929 & 0.871 & 0.791 & 0.726 & 0.674 \\
S & BAWLAS & 1.032 & 1.061 & 0.991 & 0.974 & 0.956 & 0.921 & 0.902 & 0.863 & 0.786 & 0.677 & 0.568 & 0.585 \\
Ca & ASPCAP & 0.945 & 0.865 & 0.839 & 0.814 & 0.797 & 0.775 & 0.752 & 0.732 & 0.708 & 0.675 & 0.655 & 0.652 \\
C+N & ASPCAP & 0.441 & 0.483 & 0.528 & 0.554 & 0.585 & 0.617 & 0.659 & 0.706 & 0.729 & 0.694 & 0.649 & 0.605 \\
C+N & BAWLAS & 0.673 & 0.699 & 0.687 & 0.681 & 0.689 & 0.700 & 0.722 & 0.737 & 0.732 & 0.684 & 0.558 & 0.649 \\
Na & ASPCAP & 0.394 & 0.393 & 0.441 & 0.462 & 0.489 & 0.527 & 0.560 & 0.595 & 0.568 & 0.423 & 0.228 & 0.107 \\
Na & BAWLAS & 0.732 & 0.681 & 0.579 & 0.612 & 0.601 & 0.601 & 0.586 & 0.564 & 0.532 & 0.385 & 0.144 & 0.159 \\
Al & ASPCAP & 0.893 & 0.875 & 0.902 & 0.906 & 0.925 & 0.932 & 0.929 & 0.925 & 0.917 & 0.901 & 0.851 & 0.821 \\
K & ASPCAP & 0.913 & 0.899 & 0.904 & 0.905 & 0.92 & 0.941 & 0.969 & 1.004 & 1.047 & 1.084 & 1.065 & 1.086 \\
Cr & ASPCAP & 0.405 & 0.422 & 0.428 & 0.445 & 0.455 & 0.466 & 0.482 & 0.487 & 0.466 & 0.495 & 0.549 & 0.518 \\
Fe & ASPCAP & 0.501 & 0.501 & 0.501 & 0.501 & 0.501 & 0.501 & 0.501 & 0.501 & 0.501 & 0.501 & 0.501 & 0.501 \\
Ni & ASPCAP & 0.537 & 0.561 & 0.574 & 0.582 & 0.590 & 0.597 & 0.604 & 0.606 & 0.567 & 0.517 & 0.493 & 0.448 \\
V & ASPCAP & 0.896 & 0.830 & 0.677 & 0.573 & 0.567 & 0.586 & 0.643 & 0.684 & 0.655 & 0.571 & 0.529 & 0.524 \\
V & BAWLAS & 0.996 & 0.940 & 0.914 & 0.779 & 0.719 & 0.691 & 0.697 & 0.688 & 0.628 & 0.557 & 0.527 & 0.609 \\
Mn & ASPCAP & 0.238 & 0.255 & 0.275 & 0.293 & 0.312 & 0.329 & 0.344 & 0.354 & 0.317 & 0.232 & 0.148 & 0.146 \\
Co & ASPCAP & 0.371 & 0.453 & 0.507 & 0.538 & 0.577 & 0.611 & 0.657 & 0.688 & 0.655 & 0.568 & 0.505 & 0.511 \\
Cu & BAWLAS & -- & 0.978 & 0.770 & 0.655 & 0.623 & 0.580 & 0.586 & 0.610 & 0.614 & 0.554 & 0.494 & 0.295 \\
Ce & ASPCAP & 0.500 & 0.412 & 0.396 & 0.386 & 0.374 & 0.350 & 0.351 & 0.393 & 0.431 & 0.480 & 0.562 & 0.725 \\
Ce & BAWLAS & 0.717 & 0.632 & 0.555 & 0.522 & 0.494 & 0.483 & 0.481 & 0.500 & 0.504 & 0.476 & 0.540 & -- 
\enddata
\end{deluxetable*}

\begin{deluxetable*}{cccccccccccccc}
\tablecaption{Values of $q_{\text{Ia}}^X$} \label{table:qIa}
\tablehead{\colhead{Element} & \colhead{Source} & \multicolumn{12}{c}{[Mg/H]}\\
\colhead{} & \colhead{} & \colhead{-0.7} & \colhead{-0.6} & \colhead{-0.5} & \colhead{-0.4} & \colhead{-0.3} & \colhead{-0.2} & \colhead{-0.1} & \colhead{0.0} & \colhead{0.1} & \colhead{0.2} & \colhead{0.3} & \colhead{0.4}}
\startdata
O & ASPCAP & 0.094 & 0.086 & 0.042 & 0.027 & 0.02 & 0.033 & 0.040 & 0.047 & 0.063 & 0.091 & 0.122 & 0.131 \\
O & BAWLAS & -0.034 & -0.200 & -0.122 & -0.111 & -0.079 & -0.058 & -0.032 & -0.005 & 0.023 & 0.084 & 0.058 & 0.095 \\
Mg & ASPCAP & 0.000 & 0.000 & 0.000 & 0.000 & 0.000 & 0.000 & 0.000 & 0.000 & 0.000 & 0.000 & 0.000 & 0.000 \\
Si & ASPCAP & 0.065 & 0.102 & 0.143 & 0.164 & 0.174 & 0.184 & 0.180 & 0.181 & 0.200 & 0.241 & 0.282 & 0.288 \\
S & ASPCAP & 0.207 & 0.038 & 0.012 & 0.019 & 0.022 & 0.052 & 0.065 & 0.071 & 0.096 & 0.156 & 0.192 & 0.225 \\
S & BAWLAS & 0.230 & 0.169 & 0.100 & 0.085 & 0.099 & 0.147 & 0.152 & 0.137 & 0.130 & 0.136 & 0.105 & 0.025 \\
Ca & ASPCAP & 0.199 & 0.200 & 0.23 & 0.250 & 0.267 & 0.286 & 0.285 & 0.268 & 0.264 & 0.271 & 0.272 & 0.264 \\
C+N & ASPCAP & 0.309 & 0.441 & 0.395 & 0.373 & 0.335 & 0.303 & 0.285 & 0.294 & 0.337 & 0.429 & 0.511 & 0.582 \\
C+N & BAWLAS & 0.301 & 0.380 & 0.371 & 0.342 & 0.304 & 0.271 & 0.249 & 0.263 & 0.300 & 0.389 & 0.520 & 0.437 \\
Na & ASPCAP & 0.223 & 0.508 & 0.499 & 0.483 & 0.457 & 0.407 & 0.386 & 0.405 & 0.527 & 0.808 & 1.107 & 1.329 \\
Na & BAWLAS & 0.260 & 0.700 & 0.748 & 0.589 & 0.516 & 0.432 & 0.414 & 0.436 & 0.523 & 0.768 & 1.084 & 1.185 \\
Al & ASPCAP & -0.231 & 0.225 & 0.267 & 0.249 & 0.197 & 0.135 & 0.099 & 0.075 & 0.066 & 0.073 & 0.113 & 0.141 \\
K & ASPCAP & 0.100 & 0.117 & 0.104 & 0.096 & 0.083 & 0.053 & 0.024 & -0.004 & -0.025 & 0.010 & 0.088 & 0.117 \\
Cr & ASPCAP & 0.439 & 0.459 & 0.494 & 0.503 & 0.514 & 0.511 & 0.502 & 0.513 & 0.560 & 0.562 & 0.535 & 0.578 \\
Fe & ASPCAP & 0.499 & 0.499 & 0.499 & 0.499 & 0.499 & 0.499 & 0.499 & 0.499 & 0.499 & 0.499 & 0.499 & 0.499 \\
Ni & ASPCAP & 0.363 & 0.490 & 0.459 & 0.437 & 0.412 & 0.381 & 0.374 & 0.394 & 0.466 & 0.545 & 0.590 & 0.660 \\
V & ASPCAP & 0.461 & 0.275 & 0.320 & 0.296 & 0.289 & 0.294 & 0.295 & 0.316 & 0.403 & 0.539 & 0.611 & 0.647 \\
V & BAWLAS & 0.283 & 0.440 & 0.361 & 0.367 & 0.345 & 0.325 & 0.313 & 0.312 & 0.370 & 0.448 & 0.469 & 0.381 \\
Mn & ASPCAP & 0.330 & 0.546 & 0.539 & 0.532 & 0.532 & 0.544 & 0.574 & 0.646 & 0.786 & 0.977 & 1.141 & 1.244 \\
Co & ASPCAP & 0.347 & 0.388 & 0.371 & 0.375 & 0.358 & 0.321 & 0.293 & 0.312 & 0.394 & 0.539 & 0.643 & 0.693 \\
Cu & BAWLAS & -- & 0.159 & 0.330 & 0.357 & 0.323 & 0.330 & 0.344 & 0.390 & 0.493 & 0.687 & 0.928 & 1.300 \\
Ce & ASPCAP & 0.271 & 0.544 & 0.652 & 0.738 & 0.863 & 0.955 & 0.835 & 0.607 & 0.444 & 0.252 & 0.104 & -0.073 \\
Ce & BAWLAS & 0.452 & 0.561 & 0.716 & 0.777 & 0.888 & 0.940 & 0.749 & 0.500 & 0.361 & 0.284 & 0.116 & --
\enddata
\end{deluxetable*}

\begin{deluxetable}{cc}
\tablecaption{Columns Included in Catalog Table} \label{table:catalog_cols}
\tablehead{\colhead{Label} & \colhead{Description}}
\startdata
APOGEE\_ID\tablenotemark{a} &  APOGEE object name \\
LOCATION\_ID\tablenotemark{a} & APOGEE field location ID \\
TEFF\tablenotemark{a} & Calibrated effective temperature \\
LOGG\tablenotemark{a} & Calibrated surface gravity \\
SNR\tablenotemark{a} & Estimated signal-to-noise per pixel \\
A\_CC & Best-fit $\Acc$ \\
A\_IA & Best-fit $\AIa$ \\
CHISQ & $\chi^2$ value \\
RED\_CHISQ & Reduced $\chi^2$ (Eq. \ref{eq:red_chi2}) \\
DFLAG\tablenotemark{b} & Primary/Secondary Spectrum Flag \\
X\_H\_SOURCE\_RAW\tablenotemark{c} & Raw [X/H] value \\
X\_H\_SOURCE\_ADJ & [X/H]$_\text{corr}$ value (Eq. \ref{eq:xh_corr})\\
X\_H\_SOURCE\_ERR\tablenotemark{c} & [X/Fe] uncertainty\tablenotemark{d} \\
X\_H\_SOURCE\_DEV & Two-process residual $\Delta$[X/H]\tablenotemark{e}
\enddata
\tablenotetext{a}{Taken directly from APOGEE DR17 \texttt{allStar} file.}
\tablenotetext{b}{Highest SNR spectrum indicated by \texttt{DFLAG}=0; higher numbers rank additional spectra by descending SNR.}
\tablenotetext{c}{Taken directly from APOGEE DR17 (if SOURCE=ASPCAP) or BAWLAS (if SOURCE=BAWLAS).}
\tablenotetext{d}{See Section \ref{sec:data} for details. [Fe/H] uncertainty used for X=Fe.}
\tablenotetext{e}{$\Delta\text{[X/H]} = \text{[X/H]}_{\text{corr}} - \text{[X/H]}_{\text{2proc}}$ (see Section \ref{sec:2proc}).}
\end{deluxetable}

\bibliography{2proc_bib}
\bibliographystyle{aasjournal}

\end{document}